%% file: MainArxiv.tex
\documentclass[9pt,twocolumn]{extarticle}

\usepackage[english]{babel}                             
\usepackage{amsmath,amssymb,latexsym,xfrac,nicefrac}    
\usepackage{varioref}                                   
\usepackage[square]{natbib}                                     
\renewcommand{\cite}{\citep}
\usepackage[hidelinks]{hyperref}                        
\usepackage{float,caption,subcaption}                   
\usepackage{graphicx,epic,eepic,epsfig,graphics,psfrag} 
\usepackage{algorithm,algorithmicx,algpseudocode}       
\usepackage{multirow,adjustbox}                         
\usepackage{color}                                      
\usepackage{geometry}                                   
\usepackage{authblk}                                    
\usepackage[raggedright]{titlesec}                      

\DeclareMathOperator*{\argmin}{arg\,min}

\newcommand{\email}[1]{\href{mailto:#1}{#1}}

\newcommand{\balance}{}
\newcommand{\Description}[2][]{}

\title{Synthesis of Frame Field-Aligned Multi-Laminar Structures}

\author[*,1]{Florian Cyril Stutz}
\affil[*]{Corresponding author: \email{fstu@dtu.dk}}

\author[1]{Tim Felle Olsen}
\author[1]{Jeroen Peter Groen}
\author[1]{Niels Aage}
\author[1]{Ole Sigmund}
\author[1]{Jakob Andreas B{\ae}rentzen}
\author[2]{Justin Solomon}

\affil[1]{
    Technical University of Denmark, 
    Anker Engelunds Vej 1, 2800, Kgs. Lyngby, Denmark
}
\affil[2]{
    Massachusetts Institute of Technology, 
    77 Massachusetts Ave, 02139, Cambridge, MA, United States of America
}
\date{}

\begin{document}


\twocolumn[{%
            \renewcommand\twocolumn[1][]{#1}%
            \maketitle
            \begin{center}
                \centering
                \captionsetup{type=figure}
                \includegraphics[width=\textwidth]{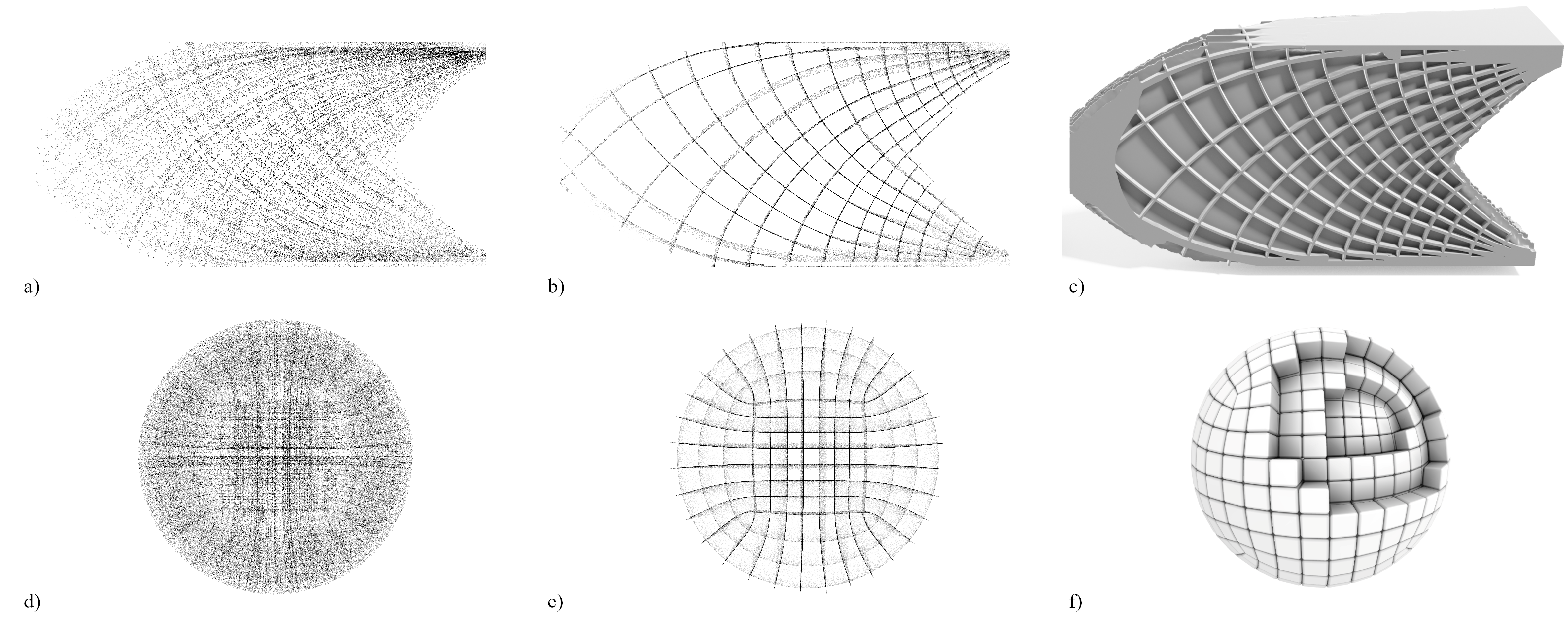}
                \captionof{figure}{
                    Given a frame field as an input we can generate a set of
                    optimal laminations aligning exactly with the field
                    orientations (Subfigures a and d). Using a novel
                    optimization energy that only needs local orientation
                    awareness, we can create a well-spaced subset of these
                    laminations (Subfigures b and e). We then proceed to create
                    near optimal, highly stiff multi-laminar structure as a
                    volumetric solid (Subfigure c) or in some cases output a
                    hexahedral mesh (Subfigure f).
                }
                \label{fig:teaser}
            \end{center}%
        }]

\input{Sections/00Abstract.tex}

\input{Sections/00Introduction.tex}

\input{Sections/00RelatedWork}
\input{Sections/00InputFields}
\input{Sections/00Subselection.tex}
\input{Sections/00StructureGeneration.tex}

\input{Sections/00Results.tex}
\input{Sections/00Conclusion.tex}
\vfill
\bibliographystyle{apalike}
\bibliography{literature}

\end{document}

%% file: Sections/00Abstract.tex
\begin{abstract}

    In the field of topology optimization, the homogenization approach has been revived as an important alternative to the established, density-based methods because it can represent the microstructural design at a much finer length-scale than the computational grid. The optimal microstructure for a single load case is an orthogonal rank-3 laminate. A rank-3 laminate can be described in terms of frame fields, which are also an important tool for mesh generation in both 2D and 3D.

    We propose a method for generating multi-laminar structures from frame fields.
    Rather than relying on integrative approaches that find a parametrization based on the frame field, we find stream surfaces, represented as point clouds aligned with frame vectors, and we solve an optimization problem to find well-spaced collections of such stream surfaces. The stream surface tracing is unaffected by the presence of singularities outside the region of interest. Neither stream surface tracing nor selecting well-spaced surface rely on combed frame fields.

    In addition to stream surface tracing and selection, we provide two methods for generating structures from stream surface collections. One of these methods produces volumetric solids by summing basis functions associated with each point of the stream surface collection. The other method reinterprets point sampled stream surfaces as a spatial twist continuum and produces a hexahedralization by dualizing a graph representing the structure.

    We demonstrate our methods on several frame fields produced using the homogenization approach for topology optimization, boundary-aligned, algebraic frame fields, and frame fields computed from closed-form expressions.

\end{abstract}

%% file: Sections/00Introduction.tex
\section{Introduction}
In recent years, \textit{topology optimization} \cite{Bendsoe2004} has emerged as an important tool in digital modeling and fabrication. By minimizing compliance, for example, topology optimization can produce mechanical structures that are stiffer than what a human designer would usually be able to achieve, using only a specified amount of material. More generally, topology optimization algorithms can directly optimize for structures, extremizing assorted quality measures of fabricated objects.

Density-based approaches for topology optimization employ a straightforward minimization over parameters that represent element-wise material density and, as such, operate directly on a volumetric shape representation. Unfortunately, large-scale topology optimization problems are very computationally demanding with this type of approach, albeit feasible in many contexts \cite{Aage2017,Baandrup2020}.

The \emph{homogenization-based} approach to topology optimization offers an alternative wherein the material is represented in terms of homogenized microstructures \cite{Bendsoe1988}. The optimal microstructure for single-load case stiffness optimization, which we also employ, is the rank-3 microstructure, a lamination system with three orthogonal lamination directions. Using the orientation and lamination thicknesses obtained during topology optimization, we can realize a physical structure from the homogenization solution at a finite length scale in a process called \emph{de-homogenization}. In practice, this two-step procedure yields finely resolved structures at a much lower computational cost than density-based methods \cite{Bendsoe2004,Pantz2008,Donders2018,GeoAllOli20,Groen2018, Groen2020}. 

Clearly, one could also choose truss-based microstructures for ho\-mog\-e\-niza\-tion-based topology optimization, resulting in a final structure consisting of trusses as done in \citet{Wu2021}. However, a truss carries load only in the direction of the truss itself, while a sheet can carry load along two directions. In practice, this means that closed wall structures are up to three times stiffer than Michell structures \cite{Sigmund2016}. Consequently, our goal is to construct closed wall structures.

The specific problem we address is the following.
Assuming the lamination orientations are given by a frame field, we seek a set of surfaces such that each surface aligns everywhere with one of the frame orientations. The surfaces should be approximately evenly-spaced, and the spacing corresponding to a choice of length scale; up to three surfaces might intersect at any point.

If we can find a 3D parametrization of the domain such that the gradients of the coordinate functions are everywhere aligned with the frame field, the surfaces are simply constant coordinate surfaces pulled back from the parametrization domain. Unfortunately, the frame field might be far from integrable, and there are few if any robust approaches that can handle such cases. While recent work modifies the frame field at the cost of structural performance to promote integrability \cite{arora2019}, we take a different route which does not require a parametrization of the domain.

We seek a set of surfaces whose local tangent planes are aligned to a frame field. We find these surfaces individually using an approach that amounts to stream surface tracing. Given a large superset of such surfaces, it is then possible to find an evenly-spaced subset by solving a binary optimization problem, which we solve efficiently through relaxation.

Our first contribution is a method that solves the aforementioned problem by tracing and selecting stream surfaces that locally align with frame fields. The tracing is discussed in Section \ref{sec:CreateStreamSurfaces} and the selection procedure in Section \ref{sec:Energies}. 

%
Given a set of stream surfaces, we further provide two methods for the synthesis of output shapes. In topology optimization, we usually need a manufacturable solid as the output.  This volumetric solid can be extracted by compositing samples of each stream surface onto a voxel grid---a procedure sometimes known as \textit{splatting}, described in Section~\ref{sec:volumetric-solids}. For frame fields that lead to reasonably isotropic families of surfaces, we also can compute a graph of the intersection points and output a combinatorial structure from which a hexahedral mesh can be obtained, as described in Section~\ref{sec:hexes}.

%% file: Sections/00RelatedWork.tex
\section{Related Work}
\label{sec:RelatedWork}



Frame fields resulting from topology optimization impose particular requirements on hexahedral mesh generation schemes. For example, the frame fields might exhibit anisotropy to an extent where one edge length deteriorates. Moreover, the rotation of the frame fields might be higher than is usual in the case of fields designed for hexahedral meshing, and this could make the fields non-integrable. These challenges suggest that existing hex-meshing or hex-dominant meshing algorithms are not suitable for such problems.

In recent years, density-based topology optimization has been used to find optimal mechanical structures in various fields. In the area of compliance minimization, giga-scale finite element models have been applied \cite{Aage2017, Baandrup2020}. While such large-scale topology optimization makes the benefits of topology optimized structures very apparent, it also relies on supercomputing hardware and is not applicable in real time, which is one of the key steps towards the goal of incorporating topology optimization in the everyday engineering design process.

Density-based topology optimization methods such as SIMP or RAMP \cite{Sigmund2013} were designed to directly produce single-scale mechanical structures. However, earlier work, specifically the groundbreaking work by \citet{Bendsoe1988}, modelled material as having an infinitesimal microstructure---as opposed to being locally characterized only by density. Materials consisting of microstructures have been shown to be computationally optimal, while circumventing the problem that density-based topology optimized structures depend on the size of the chosen finite element mesh \cite{Avellaneda1987,Francfort1995, Sigmund2013}.

The process of going from the results of homogenization based topology optimization to high-resolution structures is called \emph{de-homogenization}. It was introduced by \citet{Pantz2008}, who combined homogenization-based topology optimization with field integration, as done in quad-meshing, to de-homogenize 2D examples whose orientation fields do not contain singularities. They expanded their approach in \citet{Pantz2010} to structures with singularities of index $\pm\nicefrac{1}{2}$ lying in void regions by punching out holes around these singularities.
\citet{Groen2018} revisited this method and simplified the approach, while introducing additional parameters for more control of the de-homogenized structure. Their approach was limited to singularity-free fields and has since been ported to 3D in \citet{Groen2020}.
\citet{Donders2018} proposed a method for de-homogenizing structures in 2D with singularities of index $\pm \nicefrac{1}{2}$ without the need of punching holes based on \citet{hotz_et_al:DFU:2010:2700}.
\citet{Stutz2020} expanded the approach by \citet{Groen2018} to incorporate examples with singularities of index $\pm\nicefrac{1}{4}$.

All of the above papers indicate a strong relationship to quad-dominant meshing in 2D and hex-dominant meshing in 3D. Depending on the examples, a deterioration of the hexahedra is desirable as is the anisotropy resulting thereof. However, as shown in \cite{Groen2018, Stutz2020}, spurious singularities can occur. In 3D, orientations of the microstructures are not unique, a problem for the de-homogenization that can to a certain degree be circumvented by regularization \cite{Groen2020}.

Approaches for truss-structures have been presented for singularity-free fields in \citet{Larsen2018, arora2019} and in \citet{Wu2021} for fields containing singularities.

Field-based quad-meshing and hex-meshing is most often done by combing fields and integrating to find scalar functions with integer-jump conditions, where the combed field are differently labelled \cite{Kaelberer2007,Bommes2009,Nieser2011}.
A lot of research for field-based hex-meshing focuses on achieving pure-hex meshes \cite{Huang2011,Ray2016,Solomon2017,Palmer2019}. These methods focus on the field design part of the hex-meshing pipeline with the main goal to achieve as many hexahedral elements as possible. Thus, these methods minimize a smoothness energy while ensuring that at the surface one direction of the octahedral frame is well-aligned with the surface normal \cite{Huang2011}. As a natural effect, hex-meshes extracted from such a model tend to have minimized anisotropy and minimized deterioration of the hexahedral elements.

For de-homogenization and hex-dominant meshing of ho\-mog\-e\-niza\-tion-based topology optimization results, it is of importance to note that the fields are typically \emph{prescribed} (rather than optimized during the meshing procedure) and cannot be changed to obtain more smoothness without reducing the mechanical performance of the obtained structure \cite{Stutz2020}. Approaches like \citet{Kaelberer2007} and \citet{Nieser2011} are promising for de-homogenization but contain a major pitfall since fields arising from the homogenization method often have singularities of higher indices ($\pm\nicefrac{1}{2}$ in 2D) or have significant divergence at mechanical boundary conditions. Such higher indices imply a greater rotational speed and typically integration based methods for de-homogenization must enforce alignment to the fields with a penalization approach \cite{Groen2018,Groen2020,Stutz2020}. This penalization weight trades off structural alignment with spacing of the structural members and implicitly introduces anisotropy. If the alignment weight is chosen too small, the resulting parametrization will not align well with the underlying field as it tries to create unit-length gradients. If the alignment weight is chosen too large, the gradient of the parametrization will become zero and result in stretched out iso-contours  \cite{Stutz2020}. These problems might be mitigated by introduction of additional optimization terms, which has so far not been deeply investigated. It is important to note that anisotropy is desired and of the utmost importance for the mechanical performance.

In field-based hex-dominant meshing as done by \citet{Gao2017}, the isotropy of the desired hexahedra is a key ingredient of the algorithm. This is due to the optimization, which trades off the regularity of the hexahedra and their alignment to the underlying field. An expansion to anisotropic hex-dominant meshing might be achieved, if the desired hex-edge length was known beforehand and not only given implicitly.

\citet{Ni2018} have a promising approach to solve for vertex position of a tetrahedral mesh, which is similar to \citet{Gao2017}. The nature of the approach is aimed at producing vertices of a hex-mesh with a prescribed isotropic edge-length. Note that \citet{Gao2017} and \citet{Ni2018} create tetrahedra where the hexahedra do not align with the field, which could cost dearly in terms of mechanical performance, when used for de-homogenization, since the resulting structure would not align with the load path at all in these regions.
Recently, polycube methods have advanced the hex-meshing field, but since methods like \citet{Guo2020} and \citet{Livesu2020} do not rely on fields they are not applicable to de-homogenization.

The work of \citet{Takayama19} expanding the 2D work of \citet{Campen2012} and \cite{Campen2014} relies on user-defined (as opposed to frame field aligned) implicit surfaces as an input to guide the creation of hex-meshes. Moreover, several authors, including us, draw inspiration from the notion of the spatial twist continuum which is, essentially, the dual of a hexahedralization and was introduced by \citet{Murdoch97}.

\citet{Campen2016} create a foliation as a means of finding a bijective parametrization of a 3D shape. While there is a clear similarity between the notion of a stream surface and a transversal section of a leaf of a foliation of a 3-manifold \cite{milnor1970foliations}, their aim is to create a bijective map entailing strong conditions on the direction field whereas we take the frame field as is.

It should be mentioned that stream surfaces are often used as a visualization tool seen in fluid dynamics \cite{hultquist1992a,machado2014a}.

We strive for global surfaces or layers, which are locally well-aligned with the results from the homogenization-based topology optimization, while incorporating implicitly the anisotropy dictated by these fields and circumventing the issue of missing structural parts due to enforcing field alignment and resulting zero-gradient regions.

%% file: Sections/00InputFields.tex
\section{Sources of Frame Fields}
\label{sec:inputFields}
We are mainly motivated by fields arising from topology optimization, but one can also think of fields that would not permit generating laminations or even hexahedral meshes obtained from an integration based method. In the following, we will shortly discuss these fields and their origins. As we will demonstrate in Section \ref{sec:Results}, our algorithm can take fields from \emph{any} of these sources as input; it is designed to extract field-aligned structures while being agnostic to the source of the field.

\subsection{Topology Optimization}
\label{sec:inputFields:TopOpt}

\begin{figure}[t]
    \centering
    \begin{minipage}[t]{0.45\columnwidth}
        \centering
        \includegraphics[trim=0mm 0mm 0mm 0mm, clip, width=\textwidth,origin=c]{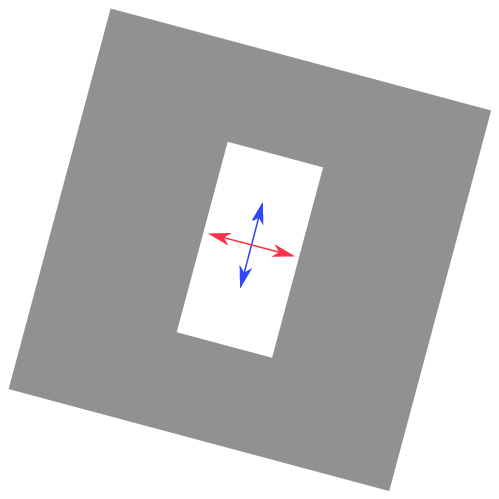}
        \subcaption{\raggedright Rectangular hole microstructure.}
        \label{fig:microstructure:UnitSquare}
    \end{minipage}
    ~
    \hfill
    ~
    \begin{minipage}[t]{0.45\columnwidth}
        \centering
        \includegraphics[trim=0mm 0mm 0mm 0mm, clip, width=\textwidth,origin=c]{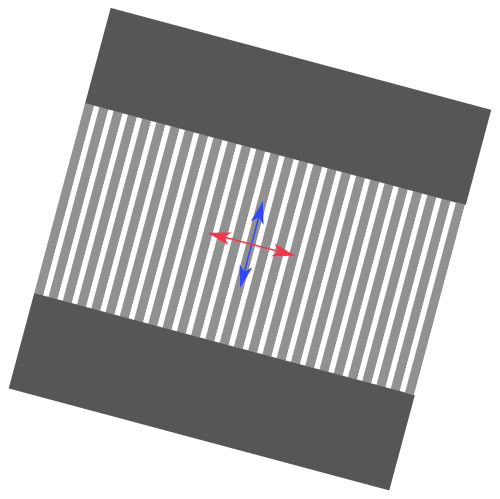}
        \subcaption{Rank-2 microstructure with orthogonal layers.}
        \label{fig:microstrucure:RankTwo}
    \end{minipage}
    \caption{Most commonly used microstructures in two dimensions. On the left the rectangular hole microstructure introduced for topology optimization in \citet{Bendsoe1988}. On the right the rank-2 microstructure with orthogonal layers first used for topology optimization in \citet{Bendsoe1989}. This rank-2 microstructure with orthogonal layers at two different length scales is known to be optimal for single strain tensor problems \cite{Avellaneda1987}.}
    \label{fig:microstructureMaterial}
\end{figure}

Homogenization-based topology optimization uses microstructures that vary their shape and orientation at an infinitesimal scale. The optimization aligns microstructures with the principal stresses implicitly \cite{Pedersen1989,Norris2005}. Two examples of microstructures are depicted in Figures \ref{fig:microstructure:UnitSquare} and \ref{fig:microstrucure:RankTwo}, the rectangular hole microstructure considered for topology optimization by \citet{Bendsoe1988} and a rank-2 material with orthogonal layers considered by \citet{Bendsoe1989}. The rectangular hole microstructure can be rotated, and the size of the hole can be changed for both directions independently. The rectangular hole microstructure is a single-scale approximation of the multiscale rank-2 material in Figure \ref{fig:microstrucure:RankTwo}. These multiscale rank-2 materials have been shown to be optimal for two-dimensional problems with a single strain tensor by \citet{Avellaneda1987}. The rank-2 microstructure can also be orientated and the relative thickness of its layers can vary independently. Note that the three dimensional equivalent to the rank-2 microstructure is called a rank-3 microstructure with orthogonal layers and is optimal for three-dimensional problems with a single strain tensor.

These infinitesimal microstructures are used in topology optimization using the theory of homogenization. By assuming periodicity at the infinitesimal microscale we can obtain effective (homogenized) properties at the macroscale. By assuming only variation of the microstructures at the macro-scale we can model a complex structure using relatively few finite elements compared to the mesh-dependent density-based topology optimization \cite{Bendsoe2004}. Now the optimizer can optimize the orientations of the microstructure and the layer-widths. We obtain a coarse representation of the locally optimal microstructure orientation and layer thicknesses as a result of the topology optimization. The orientations are described as 4-direction fields in two dimensions and as octahedral fields in three dimensions.

A crucial part of the homogenization-based topology optimization is to find the optimal rotations of the microstructures, since microstructures have a high stiffness in their principal directions but low shear. Thus regularization of the orientations during the topology optimization will influence the resulting performance of the mechanical structure since more material needs to be allocated to strongly regularized regions \cite{Stutz2020}. If regularization of the orientation fields is done after the topology optimization, either actively as discussed in \citet{arora2019} or by not enforcing high enough penalization weights for an integrative method as discussed in \citet{Groen2018} and \citet{Stutz2020}, the resulting structure will not align well to the optimal microstructure orientation. Such non-optimally aligned regions may cause a dramatic loss of performance of the structure \cite{Groen2018, Stutz2020}.
Therefore the motivation of this paper is to find structures that adhere to the local orientation of the microstructure as closely as possible outside of void or fully solid regions. This in turn introduces anisotropy between the global members of the structures. Note, however, that this anisotropy is not negatively influencing the structure from a mechanical point of view.

Singularities arise in homogenization-based topology optimization for three reasons in two dimensions \cite{Stutz2020};
\begin{itemize}
    \item Singularities in the underlying stress field will lead to singularities in the layer-normal fields since the microstructure aligns to the principal stress directions.
    \item Regularization inflicted on the layer-normal fields during the topology optimization will break up singularities with a higher index in the stress fields into multiple singularities of lower index in the layer-normal fields.
    \item In regions where the microstructure is completely solid or void, singularities can be introduced by noise. In solid regions the microstructure becomes isotropic and the optimal orientation of it becomes non-unique. In void regions the microstructure is not present and an optimal orientation of the microstructure is therefore non-existing.
\end{itemize}
\citet{Stutz2020} show that in two dimensions singularities in topology optimized layer-normal fields must occur in completely solid or void regions.
\begin{figure}[ht]
    \centering
    \begin{minipage}[t]{0.44\columnwidth}
        \centering
        \includegraphics[trim=0mm 0mm 0mm 0mm, clip, width=\linewidth,origin=c]{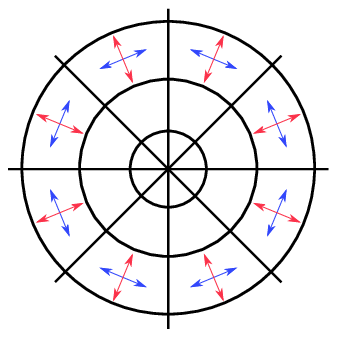}
        \subcaption{Two dimensional example of a singularity with index 1.}
        \label{fig:SingularitiesinTopOpt:Singularity2D}
    \end{minipage}
    ~
    \hspace{0.05\columnwidth} 
    ~
    \begin{minipage}[t]{0.44\columnwidth}
        \centering
        \includegraphics[trim=0mm 0mm 0mm 0mm, clip, width=\linewidth,origin=c]{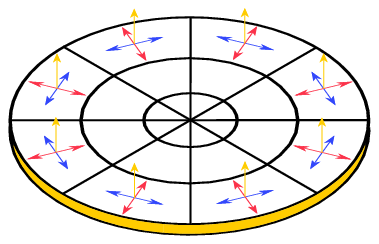}
        \subcaption{The same singularity as in Figure \ref{fig:SingularitiesinTopOpt:Singularity2D} but now embedded in an orthogonal layer (in yellow) in three dimensions.}
        \label{fig:SingularitiesinTopOpt:Singularity3D}
    \end{minipage}
    \caption{On the left we see an example of a singularity with index 1 in two dimensions. It is clear that the rotational speed increases the closer we get to the singular point. \citet{Stutz2020} have shown, that the optimizer has an incentive to put such singularities in an either fully void or fully solid region, since there would always be shear forces acting on any non-solid microstructure at the singular point. On the right we see the same field and singularity embedded into an orthogonal layer (in yellow) in three dimensions. Note how the optimizer now can choose to fill the yellow layer with material and completely ignore the red and blue field, while still creating a stiff structure. Moreover, this singularity does  not have to be in a completely solid region since the relative layer thickness of the yellow layer can be lower than 100 percent. In this case we refer to the microstructure as \emph{transversely isotropic} since the the microstructure is isotropic in one plane (the yellow one) but anisotropic perpendicular to this plane.}
    \label{fig:SingularitiesinTopOpt}
\end{figure}

Unfortunately, this observation does not hold in three dimensions. Firstly, microstructure orientations in three dimensions are non-unique due to in-plane stress; this can cause spurious singularities to appear. These singularities can be tackled with a low amount of regularization, as shown by \citet{Groen2020}.
Secondly, as seen in Figure \ref{fig:SingularitiesinTopOpt} that singularities in stress fields can occur, even when the microstructures are not completely solid.
If we consider Figure \ref{fig:SingularitiesinTopOpt:Singularity2D} we see a field describing a singularity with index 1. \citet{Stutz2020} observed that because the rotational velocity of the field increases towards infinity at the singularity, the topology optimization process fills the region around the singularity with material to account for the spinning stress-field at the singularity.
On the other hand,  when we embed the fields from Figure \ref{fig:SingularitiesinTopOpt:Singularity2D} in three dimensions, as shown in Figure \ref{fig:SingularitiesinTopOpt:Singularity3D}, the optimizer can choose to fill the newly introduced orthogonal layer with material and not assign any material to the two existing layers. Furthermore, we observe that this third layer does not have to be completely solid but can have any arbitrary layer-thickness, e.g.\ 50\%.  In this case we refer to the microstructure as being \emph{transversely isotropic} since the the microstructure is isotropic in one plane (the yellow one) but anisotropic perpendicular to this plane.
The option to cut out singularities and later on fill them with material, will inevitably lead to excessive use of material in three dimensions. The example described in Figure \ref{fig:SingularitiesinTopOpt} is, to the best of the author's knowledge, the only singularity in three dimensions that occurs outside of fully solid or entirely void regions.

The following thoughts can explain this. First, if all layer normals change direction at a location outside the void, for example, around a source, then the region would need to be filled with material by the optimizer to be made isotropic.
Second, non-zero stress directions will always be perpendicular to a layer normal, with non-zero layer thickness, meaning that stresses must always be transferred within a solid slab or plate. This will always align a stress field's singular curve with a layer normal outside of fully solid or entirely void regions. This leaves us only with fields as shown in Figure \ref{fig:SingularitiesinTopOpt:Singularity3D}, where of course, the indices of the singularities can be different.
Third, consider for a moment that the red or blue layer would be non-zero. Then their layer-normal would rotate infinitely fast at the singular curve, and thus the optimizer would fill the region completely with material to make the microstructure isotropic at the singular curve. Hence we conclude that the only singular curve not embedded into complete solid or void can be seen in Figure \ref{fig:SingularitiesinTopOpt:Singularity3D}, where the red and blue layer thicknesses are zero.

Our stream surfaces generation method can differentiate the expansion of stream surfaces near a singular region. Note how in Figure \ref{fig:SingularitiesinTopOpt:Singularity3D} the field with the yellow normal is aligned with the singular curve while having a constant normal. We can use this observation to identify which layer is traversing the singular region orthogonal to the singular curve in a computationally cheap manner and expand the corresponding stream surface through the singular region. However, in practice, we do not need to do this for topology optimized fields since we stop the expansion of stream surfaces in zero-material layers. This means that only the stream surface following the traversing layer is created.

\subsection{Boundary-Aligned Frame Fields}

Topology optimization yields frame fields as a by-product of a mechanical problem; the fields are not designed with meshability or integrability in mind.  In contrast, a number of techniques in geometry processing optimize for frame fields with the specific goal of extracting a quadrilateral or hexahedral mesh.  Since our work focuses primarily on the volumetric case, we refer the reader to \citet{Vaxman2016} for discussion of the many methods available for two-dimensional field computation, and briefly highlight representative three-dimensional methods below.

The basic goal of volumetric frame field computation is to optimize for a field of three orthogonal directions at each point in a region enclosed by a surface, with the constraint that one of the three directions aligns to the surface normal along the boundary.  This field is then used as input to methods like \citet{lyon2016hexex} and \citet{Nieser2011} to extract a mesh through parametrization.

\citet{Huang2011} originally propose a representation of orthogonal frames---later dubbed ``octahedral'' frames by \citet{Solomon2017}---that is agnostic to their labeling.  Their work extracts smooth fields by optimizing Euler angle variables, with additional constraints at the boundary; their approach was refined by \citet{Ray2016} with improved boundary constraints and optimization.  \citet{Solomon2017} propose a relaxation of \cite{Ray2016}, allowing for use of the boundary element method (BEM).  \citet{Palmer2019} provide a more complete description of the space of octahedral frames, leveraging the structure they identify to propose manifold-based optimization schemes; they also propose a related orthogonally decomposable (``odeco'') frame representation in which the directions remain orthogonal but can scale independently.

Many open questions remain regarding the singular topology of octahedral/odeco fields and its relationship to hexahedral meshing; see \citet{Liu2018} for initial results and some relevant discussion.  \citet{Corman2019} and \citet{Liu2018} propose algorithms that compute frame fields with prescribed singular structures.

\subsection{Closed-Form Frame Fields}
\label{sec:InputFields:ClosedForm}
A closed-form frame field is a field where the orientations of the frames can be found using a closed-form mathematical expression instead of being found using optimization or by solving a system of equations. In this paper, we consider a field describing a cylinder, much like the field illustrated in Figure \ref{fig:SingularitiesinTopOpt}, where there is a single singular curve in the center of the cylinder. Suppose one tries to extract well-aligned hexahedra from such a field using integrative methods. In that case, one will be challenged due to the high anisotropy of the hexahedral elements, which can not be treated by methods, that were designed to create isotropic hexahedra \cite{Stutz2020}. Note that the edge length of hexahedra will ultimately deteriorate towards the singular curve with such a cylinder field. A stream surface based approach can be designed to expand through singular curves for the cylinder's near-constant field (as discussed earlier in Section \ref{sec:inputFields:TopOpt}), while creating highly anisotropic hexahedra in the remaining domain. Extending this example, we also run our algorithm on a non-integrable field describing a helicoid. Again, this produces highly anisotropic hexahedra matching the spiral shape of the input field.

%% file: Sections/00Subselection.tex
\section{Computing Collections of Stream Surfaces}
\label{sec:ComputingSubselection}
The overarching idea of our method is to compute a large set of surfaces,  $\mathcal{S}$, which align with the frame field and then find a well-spaced selection of these, $\mathcal{S}_{opt}$, to get a representation of the multi-laminar structure that we seek. In this section, we discuss how we find and select these aligned surfaces using stream surface tracing. In the next section, Section \ref{sec:OutputGeneration}, we will discuss how the final output is computed from this representation.

In engineering, a \textit{streamline} is simply a curve that is everywhere tangential to a vector field \cite{hultquist1992a}. A stream surface is simply the generalization to 3D, i.e.\ a surface whose normal is everywhere aligned with one of the vectors of the input frame field.

We cannot rely on the frame field being combed, and hence, we do not have consistent labeling of the vectors in the frame. This is handled by simply finding the frame vector best aligned with the estimated normal of the next point that we compute when expanding a stream surface. It is also worth noting that we generally wish to stop stream surface tracing when the stream surface would otherwise exit a given bounding shape. Thus, we assume a known mask or layer thickness in the following.





\subsection{Tracing Stream Surfaces}
\label{sec:CreateStreamSurfaces}
%
%
We start by tracing stream surfaces to create the set $\mathcal{S}$. The stream surfaces are traced independently, starting from random seed points in the domain. Rather than constructing a surface connectivity, we compute a point cloud for each stream surface. The points are placed using a method similar to the technique for \emph{Poisson Disk Sampling} (PDS) sampling introduced by \citet{Bridson2007}, except that our points are placed on a surface in 3D and are not filling the entire 3D domain.

\begin{figure}[ht]
    \includegraphics[width=0.8\columnwidth]{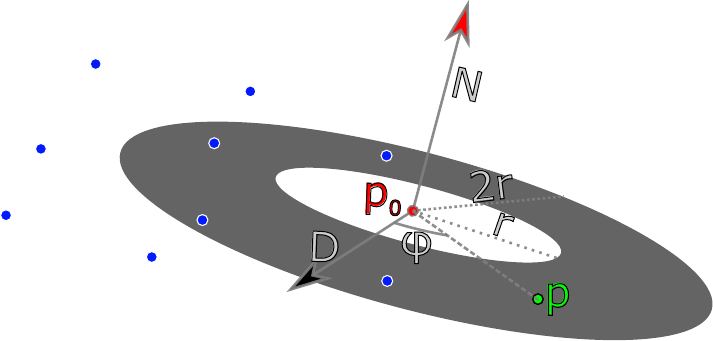}
    \Description[Generation of new points.]{This figure depicts how a new sample
        point is generated inside the annulus centered at the current point p
        zero oriented perpendicular to the surface normal N.}
    \caption{Outline of how a new point $\mathbf p$ is generated at a random position inside the
        annulus around $\mathbf p_0$ oriented perpendicularly to the desired surface
        normal $\mathbf N$.}
    \label{fig:poin_gen}
\end{figure}

We initialize each surface with a single seed point $\mathbf p_0$ and with two of the three frame vectors at $\mathbf p_0$. The first vector is our desired surface normal $\mathbf N$ at the seed point, and the second vector is perpendicular to $\mathbf N$ and describes our rotational origin $\mathbf D$. New points are now generated in an annulus centered on the seed point and perpendicular to the surface normal. Uniformly distributed random variables control the rotation angle from $\mathbf D$ and distance from $\mathbf p_0$. The annulus has an inner radius of $r$, which is the minimum distance allowed between points. The outer radius is set to $2r$ in accordance with Bridson's algorithm \cite{Bridson2007}. Each time we generate a new point, we check if it is too close to any previously generated points of the stream surface, using a lookup grid for efficiency. This generation process is visualized in \autoref{fig:poin_gen}. When a new point is accepted, it is added to a queue of points used to further expand the surface.

To mitigate drift, we employ the fourth order Runge-Kutta method
\cite{Chapra2012}. Starting from a previously determined point, $\mathbf p_0$,
with normal $\mathbf N$, we search in direction $\mathbf d_0$ with step length
$\Delta$. We need a parallel transform operator $\mathcal{P} : \mathbb{R}^3
    \rightarrow \mathbb{R}^3$ to transform the initial direction $\mathbf d_0$ onto
the tangent plane estimated at a given point $\mathbf x$ with a normal defined
by the field. The RK4 method combines partial steps through a weighted sum, to
estimate the new point. The full update can be described by,
\begin{equation*}
    \begin{aligned}
        \mathbf k_1 & = \Delta \cdot \mathcal{P}(\mathbf p_0                         ) \enspace ,                      \\
        \mathbf k_2 & = \Delta \cdot \mathcal{P}(\mathbf p_0 + \frac{\mathbf k_1}{2} ) \enspace ,                      \\
        \mathbf k_3 & = \Delta \cdot \mathcal{P}(\mathbf p_0 + \frac{\mathbf k_2}{2} ) \enspace ,                      \\
        \mathbf k_4 & = \Delta \cdot \mathcal{P}(\mathbf p_0 + \mathbf k_3           ) \enspace ,                      \\
        \mathbf p_n & = \mathbf p_0 + \frac{1}{6}(\mathbf k_1 + 2\mathbf k_2 + 2\mathbf k_3 + \mathbf k_4 ) \enspace .
    \end{aligned}
\end{equation*}

While this method is relatively precise, some drift is still unavoidable. To
improve precision, we compute the position of $p_n$ had $p_0$ instead been any
point inside the sphere with radius $2r$ centered at $p_n$. These new estimates
are avearaged to produce the new point $p$.

We discard a point if it is too close to a neighbor, distance $<r$. This could
still allow for spiralling surfaces, therefore we also look at all neighbours
within $3r$. If any of these neighbours, when projected onto the tangent plane
defined by $p$ and the associated normal, are closer to $p$ than $r$ we also
discard the point. We also do not expand stream surfaces into regions where the
corresponding layer has a layer-thickness of zero. No material would be assigned
in these regions by the splatting method described in
\autoref{sec:volumetric-solids}. We also do not expand into fully solid regions,
since these areas will be filled with material anyways by the splatting
procedure. Note that these two last cases are why we do not need to actively
handle singular curves in fully solid or entirely void regions.

\begin{figure}[ht]
    \centering
    \includegraphics[width=0.49\columnwidth]{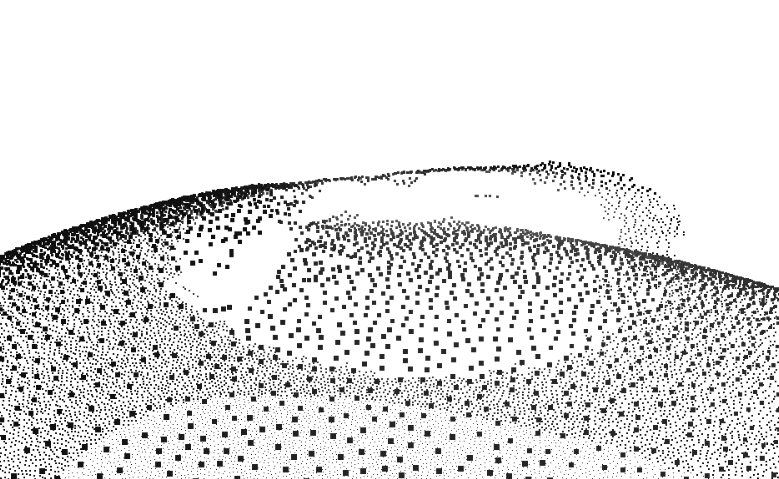}
    \includegraphics[width=0.49\columnwidth]{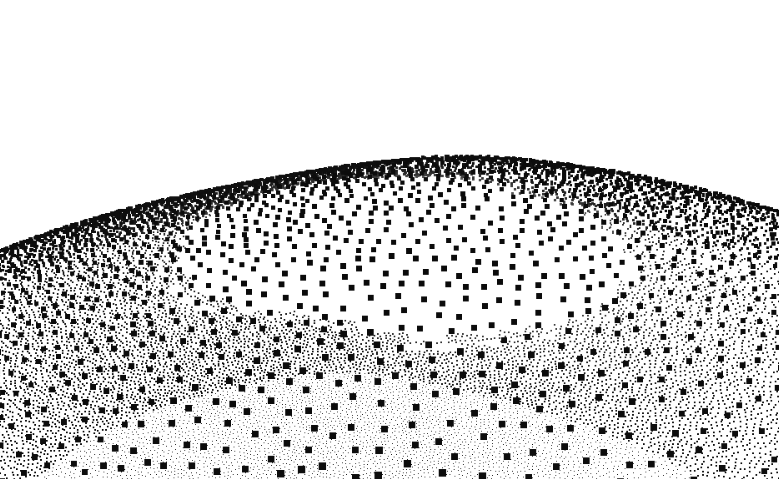}
    \caption{Here we see the effect of drifting. Small deviations in point
        position and interpolation of the field over long distances lead to a
        crack in the surface. To the right, we show the resulting surface after
        smoothing.}
    \label{fig:torsion_crack}
\end{figure}

Unfortunately, the computed stream surfaces are not entirely independent of the
starting point. The drifting can be problematic when the front of the stream
surface meets itself having traced around a round object (see the torsion sphere
example in \autoref{sec:Results:VolumetricStructures} and closeup in
\autoref{fig:torsion_crack}). This can lead to seams in the surface where it
does not entirely close up. To mitigate this problem, we post-process the surface
by recomputing the positions of all points from their neighbors using the scheme
described above. This will strengthen alignment to the field, since we use the
field information from all directions instead of only behind the expanding
front. Pseudocode for the above algorithm is given in
Algorithm~\ref{alg:surface_creation}.

We now have a method that allows us to trace stream surfaces in our input fields.
%
%
\begin{algorithm}[ht]
    \caption{Stream surface creation}
    \label{alg:surface_creation}
    \begin{flushleft}
        \textit{Inputs}: Frame field, lamination thicknesses, list of seed points, list of probe points.
    \end{flushleft}


    \begin{algorithmic}[1]
        \For{each seed point $p_s$,}
        \State Initialize queue $Q$ with $p_s$ and PDS grid with desired radius.
        \While{$Q$ is not empty,}
        \State Set $\mathbf p_0$ to front of $Q$.
        \State Compute 30 new points $p_i$ using weighted RK4.
        \If{$p_i$ is a valid point,}
        \State Save $p_i$ to point-cloud $S_s$, PDS grid and $Q$.
        \EndIf
        \EndWhile

        \For{every point $\mathbf p$ in $S_s$,}
        \State Re-estimate $\mathbf p$ using weighted RK4.
        \EndFor
        \State Save $S_s$ to the set of surfaces $\mathcal{S}$.
        \EndFor
    \end{algorithmic}

\end{algorithm}
\subsection{Singularities}
\label{sec:singularity}
\begin{figure}[ht]
    \includegraphics[width=0.95\columnwidth]{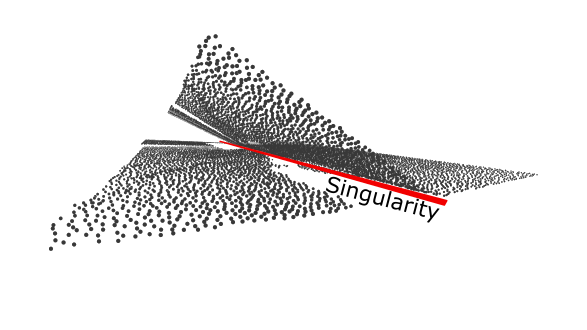}
    \Description{A surface which collides with a singularity marked with red
        splits into multiple sheets on the opposite side.}
    \caption{
        Here we see a stream surface that has hit a singularity during the
        expansion through the domain. We see the surface fork into several sheets,
        which is highly undesirable and is caused by numerical imperfections when
        generating the stream surface and rapid changes in the frame field.}
    \label{fig:singularity}
\end{figure}

When tracing stream surfaces, we will inevitably expand into regions containing singularities. In 3D, singularities are curves along which the frame field is not defined. For a more complete discussion of the notion, we refer the reader to the comprehensive overview by \citet{Vaxman2016}. In this context, singularities are challenging because the frame field rotates quickly in their vicinity. In particular, singularities of index $\pm 1$ are challenging since the frame makes a complete turn around these. These rapid rotations mean that we cannot reliably continue tracing the stream surface when it is close to a singularity. It would often lead to the stream surface effectively splitting into two or more parts (which we call "forking") due to local variations in the amount of rotation. We illustrate a forking stream surface in \autoref{fig:singularity}.


Forking surfaces can be detrimental to the quality of our results. If a surface forks, it is challenging to obtain a uniform spacing between surfaces in the selection part of our approach in \autoref{sec:ComputingSubselection:subsec:Optimization}. 
Therefore, it is essential to have a way to handle these problematic regions. We choose to stop tracing surfaces in areas close to singularities. This approach will produce holes in our models (see for example \autoref{fig:helix-hexes}). However, we preserve the overall quality of the output. As explained in Section \ref{sec:inputFields:TopOpt} cutting out the singular curve in topology optimized fields does not cause problems. Recall that for singularities outside of complete solid or void only the case in Figure \ref{fig:SingularitiesinTopOpt:Singularity3D} occurs, where we can track the layer traversing the singular curve perpendicular by its nearly constant layer normal.


Identifying singularities is a common problem in meshing and a wide range of
other fields. We are using an approach adapted from the combing approach in
\citet{Groen2020}.  The rotational energy at a voxel is computed by examining the rotation needed to align the frame in a voxel to the frames in the neighboring voxels. The average rotation is saved since we are looking for spikes in the energy. Once we have computed the energy value at every voxel, we compute some simple statistics to identify singular curves.

We then prevent stream surfaces from entering regions close to these singular curves and now have everything that we need to create the set of stream surfaces $\mathcal{S}$


\subsection{Energy for an Optimization-Based Subselection Approach}
\label{sec:Energies}
We will now take the set of surfaces $\mathcal{S}$ that we have created in the previous sections and continue with finding a well-spaced subset $\mathcal{S}_{opt}$. We will compute $\mathcal{S}_{opt}$ by optimizing over binary variables $w$ that will be assigned to the stream surfaces. However, before we can define our optimization problem, we need to define the contribution of each stream surfaces to the optimization energy. For simplicity and consistency with the figures, we will describe this procedure in two dimensions. The algorithm works the same in three dimensions, and we will explain essential details for the implementation inline on an ongoing basis.

First let $\gamma$ denote the desired average spacing in the set $\mathcal{S}_{opt}$. As an aid, we define the projection of a point $\mathbf{x} \in \mathbb{R}^{2}$ onto a streamline $S \in \mathbb{R}^2$ as $\mathbf{x}_{p} = \argmin_{x_s \in S} ||\mathbf{x} - \mathbf{x_s}||$. We can then define an energy for the streamline $S$ by
\begin{equation}
    \label{eq:EnergyOneDimensional}
    \begin{aligned}
         & \bar{E}_{S}: \mathbb{R}^{2} \rightarrow \{0, 1\} \enspace , \\
         & \bar{E}_{S}(\mathbf{x}) =
        \begin{cases}
            1, & \text{if } ||\mathbf{x} - \mathbf{x_p}|| \leq \gamma, \\
            0, & \text{else.}
        \end{cases}
    \end{aligned}
\end{equation}

\begin{figure*}[th]
    \centering
    \begin{minipage}[t]{0.48\linewidth}
        \centering
        \includegraphics[trim=0mm 0mm 0mm 0mm, clip, width=\linewidth,origin=c]{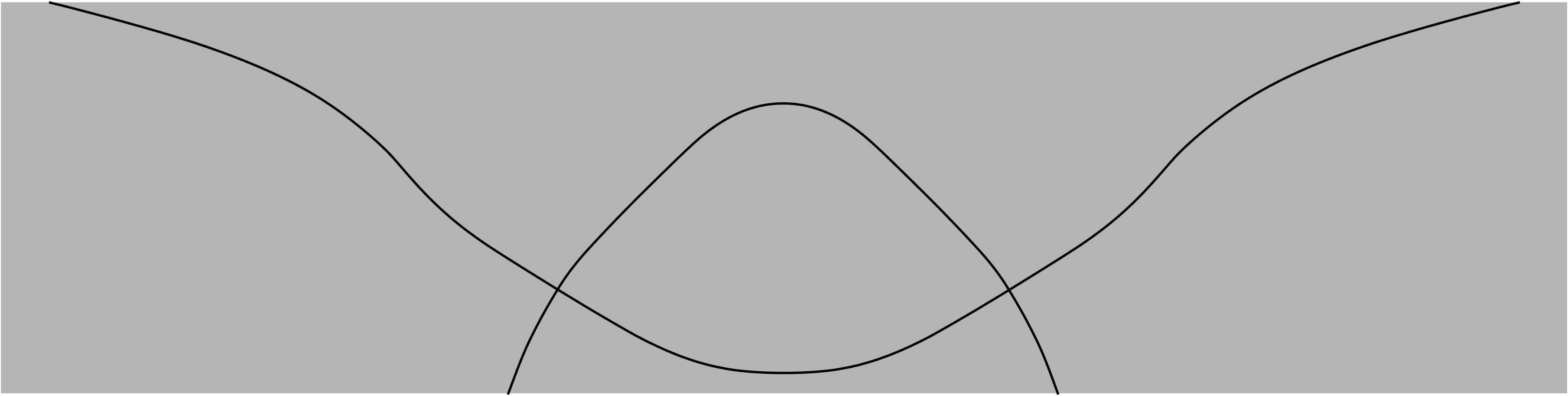}
        \subcaption{Two streamlines following two different orthogonal fields.}
        \label{fig:Energies:StreamlinesAndSingleEnergy:Streamlines}
    \end{minipage}
    ~
    \hfill
    ~
    \begin{minipage}[t]{0.48\linewidth}
        \centering
        \includegraphics[trim=0mm 0mm 0mm 0mm, clip, width=\linewidth,origin=c]{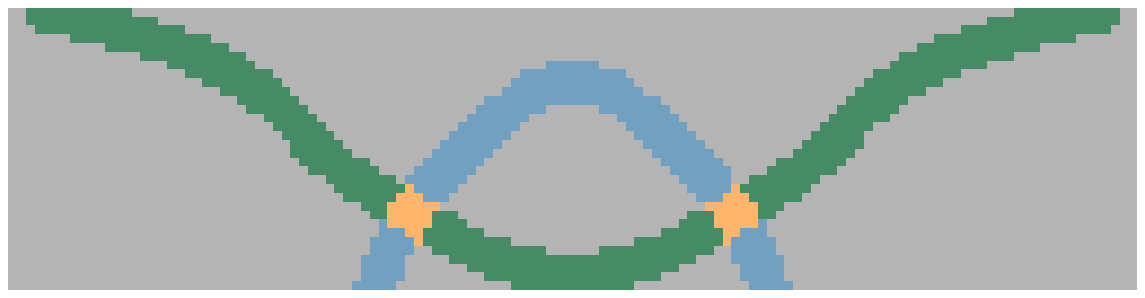}
        \subcaption{Sum of energies $\bar{E}_{S}$ for the streamlines in Figure \ref{fig:Energies:StreamlinesAndSingleEnergy:Streamlines}. Green and blue elements have an energy value of one, orange elements have a value of two.}
        \label{fig:Energies:StreamlinesAndSingleEnergy:EnergyUncombed}
    \end{minipage}
    \caption{On the left, we see two streamlines following orthogonal field directions. On the right, we see the sum of the energies $\bar{E}_{S}$ from Equation \ref{eq:EnergyOneDimensional} for the two streamlines. Here the contributions of the streamlines are colored in blue and green in regions with value one. The orange highlighted regions are elements where both streamlines create an energy response and subsequently the summed value equals two. An optimizer would try to minimize the amount of these orange elements since it tries to minimize overlapping streamline-energies. This version of the energy is blind for the fact that the two streamlines follow different fields. In order to be able to space out both family of streamlines correctly, we need to split the energy as shown in Equations \ref{eq:EnergyTwoDimensional} and Figures \ref{fig:Energies:Schema} and \ref{fig:Energies:SplitEnergies}.}
    \label{fig:Energies:StreamlinesAndSingleEnergy}
\end{figure*}

\begin{figure*}[ht]
    \centering
    \begin{minipage}[t]{0.30\linewidth}
        \centering
        \includegraphics[ trim= 10mm 35mm 10mm 32mm, clip, width=\linewidth,origin=c]{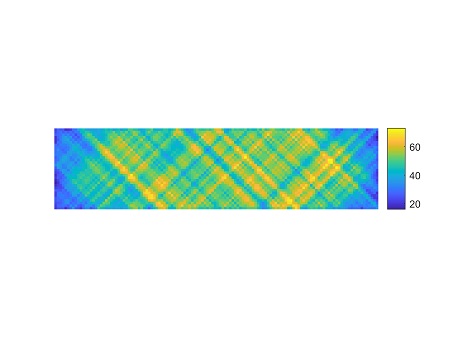}
        \subcaption{The sum of the energies $E_{S}$ of all streamlines provided to the optimizer.}
        \label{fig:SubSelection:Energy1:AllEnergies}
    \end{minipage}
    ~
    \hfill
    ~
    \begin{minipage}[t]{0.30\linewidth}
        \centering
        \includegraphics[trim= 10mm 35mm 10mm 32mm, clip, width=\linewidth,origin=c]{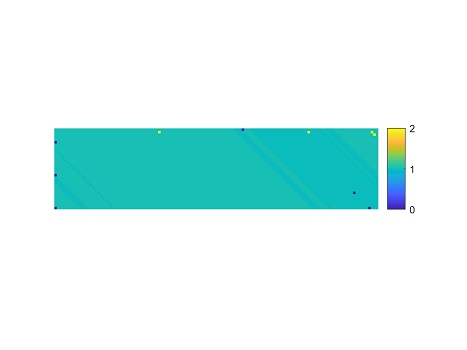}
        \subcaption{The sum of the energies $E_{S}$ of the streamlines selected by the optimizer.}
        \label{fig:SubSelection:Energy1:SelectedEnergies}
    \end{minipage}
    ~
    \hfill
    ~
    \begin{minipage}[t]{0.30\linewidth}
        \centering
        \includegraphics[ trim=0mm 15mm 0mm 10mm,clip, width=\linewidth,origin=c]{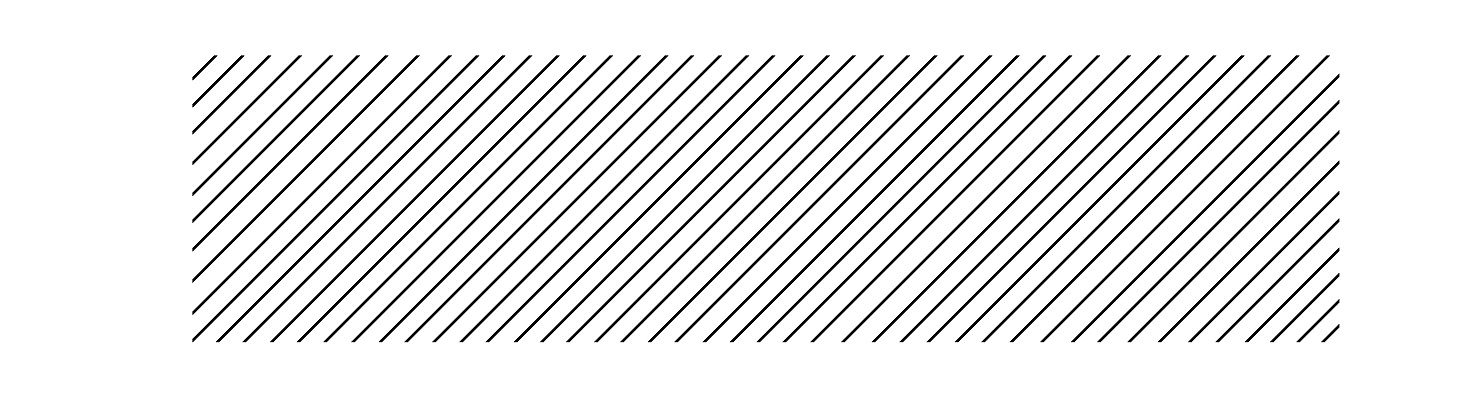}
        \subcaption{The curves selected by the optimizer.}
        \label{fig:SubSelection:Energy1:SelectedCurves}
    \end{minipage}
    \caption{Optimization results using the energies defined in Equation \ref{eq:EnergyOneDimensional}.}
    \label{fig:SubSelection:Energy1}
\end{figure*}

\begin{figure*}[ht]
    \centering
    \begin{minipage}[t]{0.30\linewidth}
        \centering
        \includegraphics[trim=12mm 10mm 12mm 10mm, clip, width=\linewidth,origin=c]{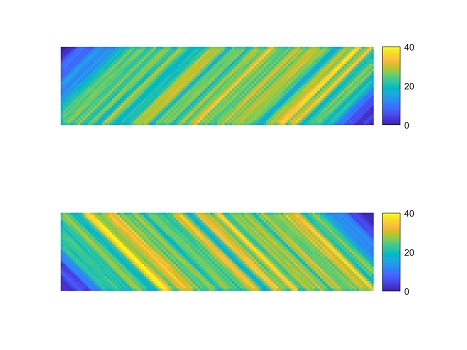}
        \subcaption{The sum of the energies $E_{S_1}$ and $E_{S_2}$ of all streamlines provided to the optimizer.}
        \label{fig:SubSelection:Energy2:AllEnergies}
    \end{minipage}
    ~
    \hfill
    ~
    \begin{minipage}[t]{0.30\linewidth}
        \centering
        \includegraphics[trim=12mm 10mm 12mm 10mm, clip, width=\linewidth,origin=c]{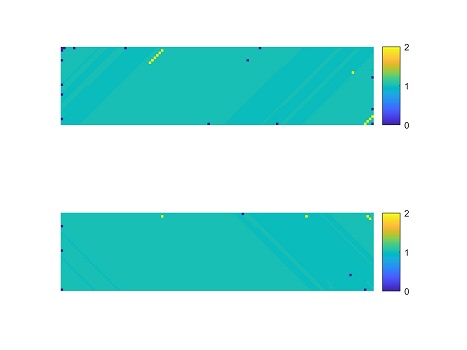}
        \subcaption{The sum of the energies $E_{S_1}$ and $E_{S_2}$ of the streamlines selected by the optimizer.}
        \label{fig:SubSelection:Energy2:SelectedEnergies}
    \end{minipage}
    ~
    \hfill
    ~
    \begin{minipage}[t]{0.30\linewidth}
        \centering
        \includegraphics[trim=  0mm 15mm 0mm 10mm, clip, width=\linewidth,origin=c]{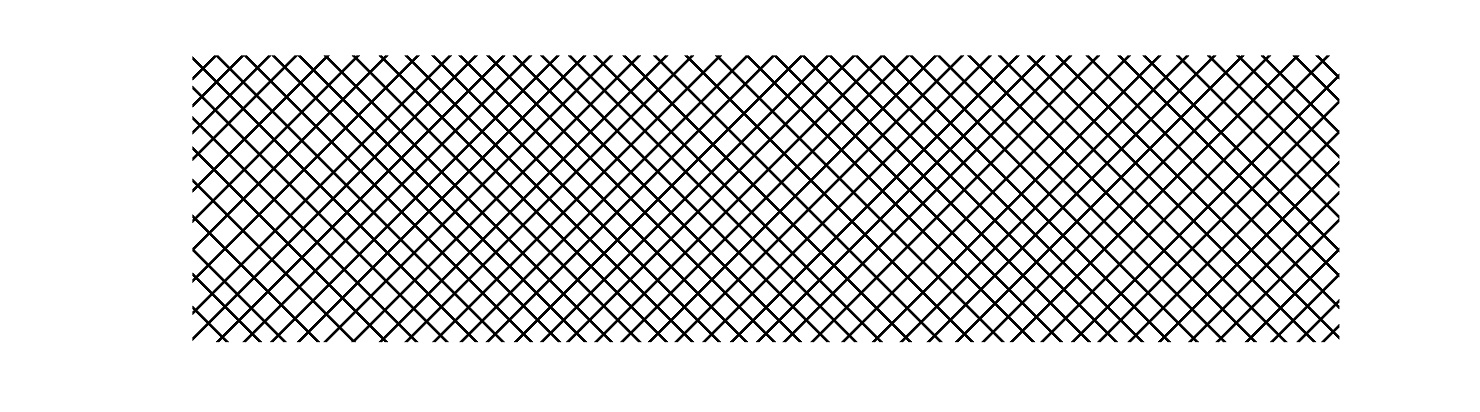}
        \subcaption{The curves selected by the optimizer.}
        \label{fig:SubSelection:Energy2:SelectedCurves}
    \end{minipage}
    \caption{Optimization results using the energies defined in Equation \ref{eq:EnergyTwoDimensional}.}
    \label{fig:SubSelection:Energy2}
\end{figure*}

\begin{figure*}[ht]
    \centering
    \begin{minipage}[t]{0.48\linewidth}
        \centering
        \includegraphics[trim=0mm 0mm 0mm 0mm, clip, width=\linewidth,origin=c]{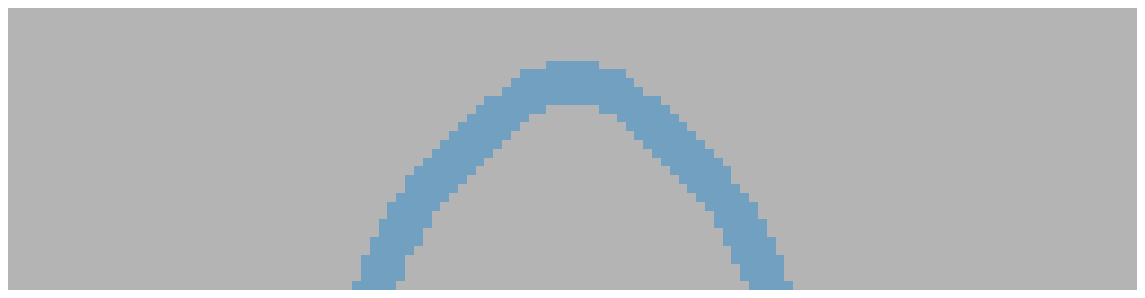}
        \subcaption{Sum of energies $E_{S_1}$ for a combed field.}
        \label{fig:Energies:EnergyCombed1}
    \end{minipage}
    ~
    \hfill
    ~
    \begin{minipage}[t]{0.48\linewidth}
        \centering
        \includegraphics[trim=0mm 0mm 0mm 0mm, clip, width=\linewidth,origin=c]{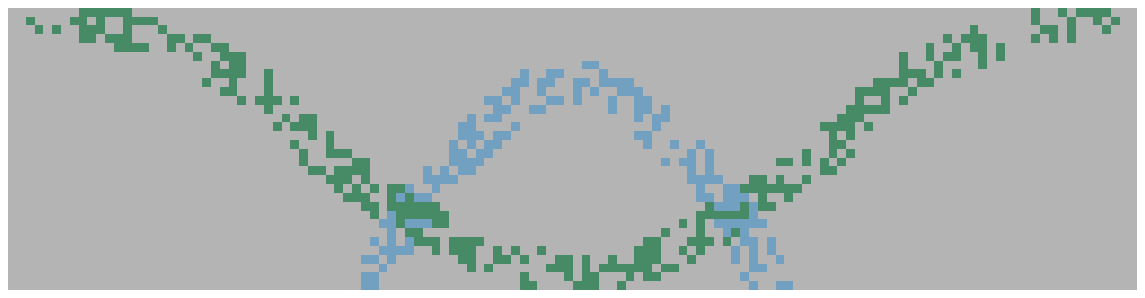}
        \subcaption{Sum of energies $E_{S_1}$ for an uncombed field.}
        \label{fig:Energies:EnergyUnCombed1}
    \end{minipage}
    \vfill{}
    \begin{minipage}[t]{0.48\linewidth}
        \centering
        \includegraphics[trim=0mm 0mm 0mm 0mm, clip, width=\linewidth,origin=c]{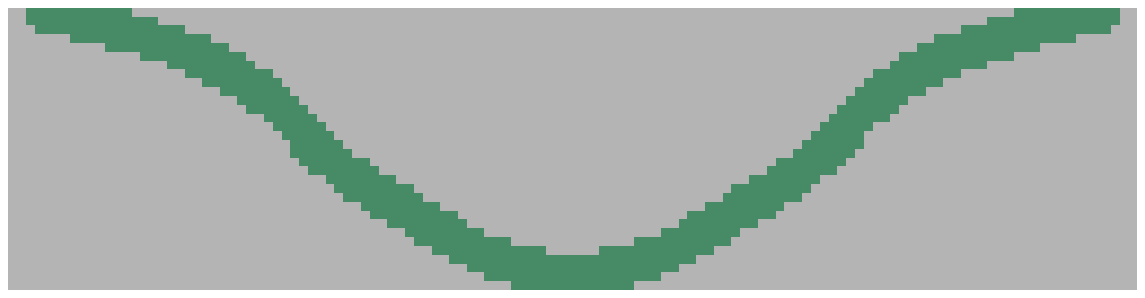}
        \subcaption{Sum of energies $E_{S_2}$ for a combed field.}
        \label{fig:Energies:EnergyCombed2}
    \end{minipage}
    ~
    \hfill
    ~
    \begin{minipage}[t]{0.48\linewidth}
        \centering
        \includegraphics[trim=0mm 0mm 0mm 0mm, clip, width=\linewidth,origin=c]{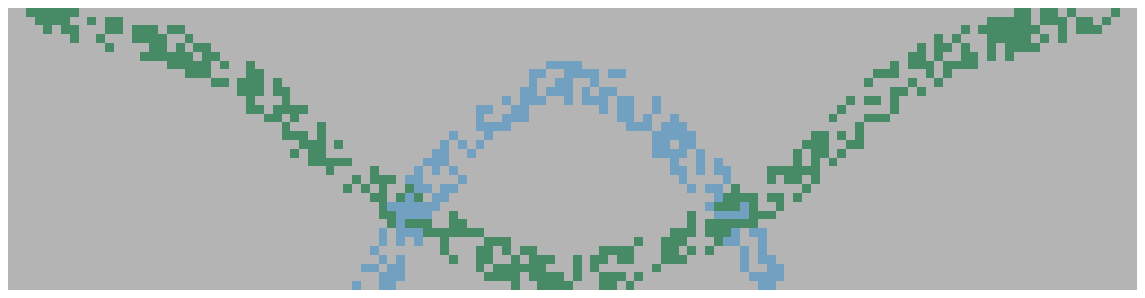}
        \subcaption{Sum of energies $E_{S_2}$ for an uncombed field.}
        \label{fig:Energies:EnergyUncombed2}
    \end{minipage}
    \caption{Sum of split energies from Equation \ref{eq:EnergyTwoDimensional} for the streamlines depicted in Figure \ref{fig:Energies:StreamlinesAndSingleEnergy:Streamlines}. The sums of the energies are shown once when a combed version of the underlying field is used and once when an uncombed version is used. Note how in a combed field the streamlines are separated into the energy that follows the same field label as the streamlines. In an uncombed field, the contributions of the streamlines split to both energies. However, it is clear that the contributions of a streamline to the two energies form a disjoint union. Regions with value two (highlighted in orange in Figure \ref{fig:Energies:StreamlinesAndSingleEnergy}) do no longer exist.}
    \label{fig:Energies:SplitEnergies}
\end{figure*}

This energy is shown in \autoref{fig:Energies:StreamlinesAndSingleEnergy:EnergyUncombed} for the two streamlines following orthogonal field directions in \autoref{fig:Energies:StreamlinesAndSingleEnergy:Streamlines}.
For our application to 4-direction fields, we need to distinguish between the two orthogonal field directions locally. Therefore, we choose for every $\mathbf{x} \in \Omega$, two orthogonal directions from the 4-direction field at random and assign them to 2-direction fields $f_1$ and $f_2$. This assignment of the orthogonal directions to $f_1$ and $f_2$ allows us to define a function $S_{dir} (\mathbf{x}_{s}) = \{1,2\}$ that indicates for every point $\mathbf{x}_s \in S$ if the streamline follows the local label of field $f_1$ or field $f_2$. In three dimensions, we use the normal of the stream surface as a field identifier.
We now split the energy for every streamline into two parts:
\begin{equation}
    \label{eq:EnergyTwoDimensional}
    \begin{aligned}
         & E_{S}: \mathbb{R}^{2} \rightarrow \{0, 1\} \times \{0, 1\} \enspace , \\
         & E_{S} = (E_{S_1}, E_{S_2}) \enspace ,
    \end{aligned}
\end{equation}
\begin{figure}[ht]
    \centering
    \includegraphics[width=\columnwidth, origin=c, trim=0mm 0mm 0mm 0mm, clip]{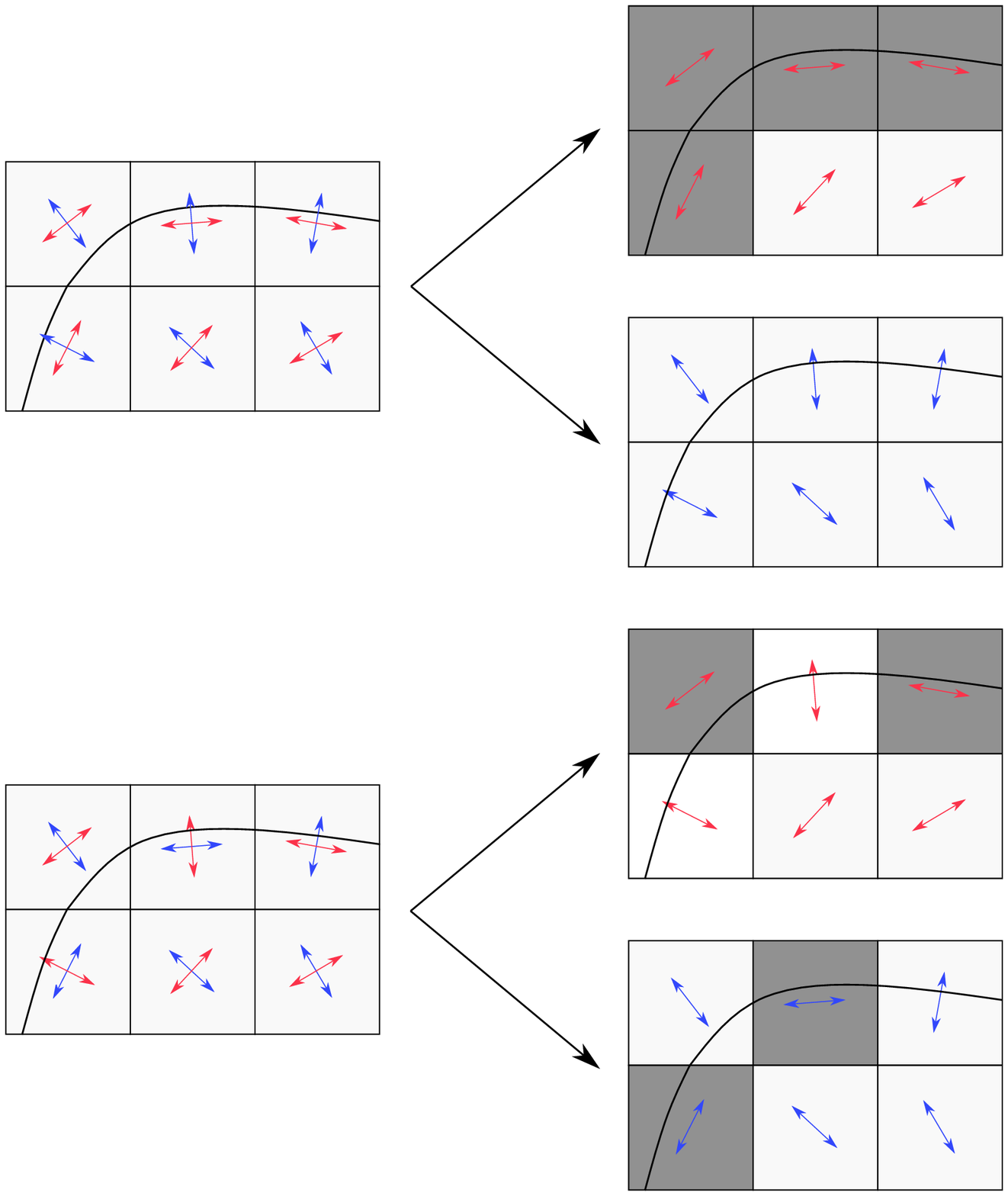}
    \caption{Example of the energies created from a streamline traced in the same field but differently labeled. On the left, we see a streamline traced in a combed (above) and uncombed (below) version of a 4-direction field. On the right, we see the corresponding energies for label "red" and label "blue", which are local labels. Note how the streamline activates only one of the two energies for each element, as indicated by the gray coloring.}
    \label{fig:Energies:Schema}
\end{figure}
where the split energies $E_{S_1}$ and $E_{S_2}$ are defined as,
\begin{equation}
    \label{eq:SplitEnergies}
    \begin{aligned}
         & E_{S_1}(\mathbf{x}) =
        \begin{cases}
            1, & \text{if} ~ S_{dir}(\mathbf{x}_{p}) = 1 \wedge ||\mathbf{x} - \mathbf{x}_{p}|| \leq \gamma \enspace , \\
            0, & \text{else,}
        \end{cases} \\
         & E_{S_2}(\mathbf{x}) =
        \begin{cases}
            1, & \text{if} ~ S_{dir}(\mathbf{x}_{p}) = 2 \wedge ||\mathbf{x} - \mathbf{x}_{p}|| \leq \gamma \enspace , \\
            0, & \text{else.}
        \end{cases}
    \end{aligned}
\end{equation}
Note that there is no need for consistency of the field labels $f_{1}$ or $f_{2}$ in a neighborhood, i.e.\ no combing is needed, as shown in Figure \ref{fig:Energies:Schema}. This makes the energies very simple to implement and the approach very robust. The split energies from Equation \ref{eq:SplitEnergies} can be seen in Figure \ref{fig:Energies:SplitEnergies}.
Having defined the energy we can now formulate a binary optimization problem,
\begin{equation}
    \label{eq:SubselectionOptimization}
    \begin{aligned}
        \underset{\mathbf{w} \in \{0,1\}^{n_{\mathcal{S}}}}{\text{minimize}}
        \int_{\Omega} \left| \sum_{i = 1}^{n_{S}} w(i) E_S(\mathbf{x}) - (1,1) \right| \quad \mathrm{d} \Omega \enspace ,
    \end{aligned}
\end{equation}
where we refer to the optimization variables $w_{i}$ as weights and $n_{S} = \vert \mathcal{S} \vert$. If we were to use the energies from Equation \ref{eq:EnergyOneDimensional} the selected streamlines would all follow the same lamination direction, since the optimizer would penalize crossing streamlines. This can be seen in Figure \ref{fig:SubSelection:Energy1}. If we use the same set of streamlines but use the energies defined in Equation \ref{eq:EnergyTwoDimensional} for the optimization we obtain both lamination as can be seen in Figure \ref{fig:SubSelection:Energy2}

\begin{figure*}[ht]
    \centering
    \begin{minipage}[t]{0.45\linewidth}
        \centering
        \includegraphics[trim =12mm 10mm 12mm 10mm, clip, width=\linewidth,origin=c]{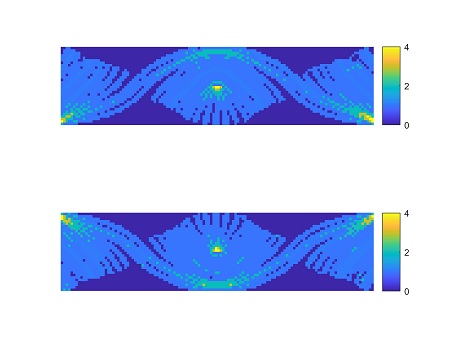}
        \subcaption{The sum of the energies $E_{S_1}$ and $E_{S_2}$ when using an L1 norm for selecting streamlines.}
        \label{fig:SubSelection:L1L2Norm:L1Norm:AllEnergies}
    \end{minipage}
    ~
    \hfill
    ~
    \begin{minipage}[t]{0.45\linewidth}
        \centering
        \includegraphics[trim = 12mm 10mm 12mm 10mm, clip, width=\linewidth,origin=c]{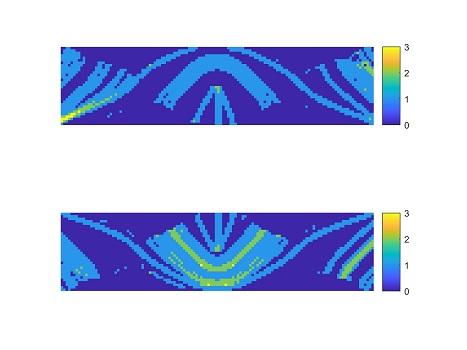}
        \subcaption{The sum of the energies $E_{S_1}$ and $E_{S_2}$ when using an L2 norm for selecting streamlines.}
        \label{fig:SubSelection:L1L2Norm:L2Norm:SelectedEnergies}
    \end{minipage}
    \vfill{}
    \begin{minipage}[t]{0.45\linewidth}
        \centering
        \includegraphics[trim=37mm 10mm 12mm 10mm, clip, width=\linewidth,origin=c]{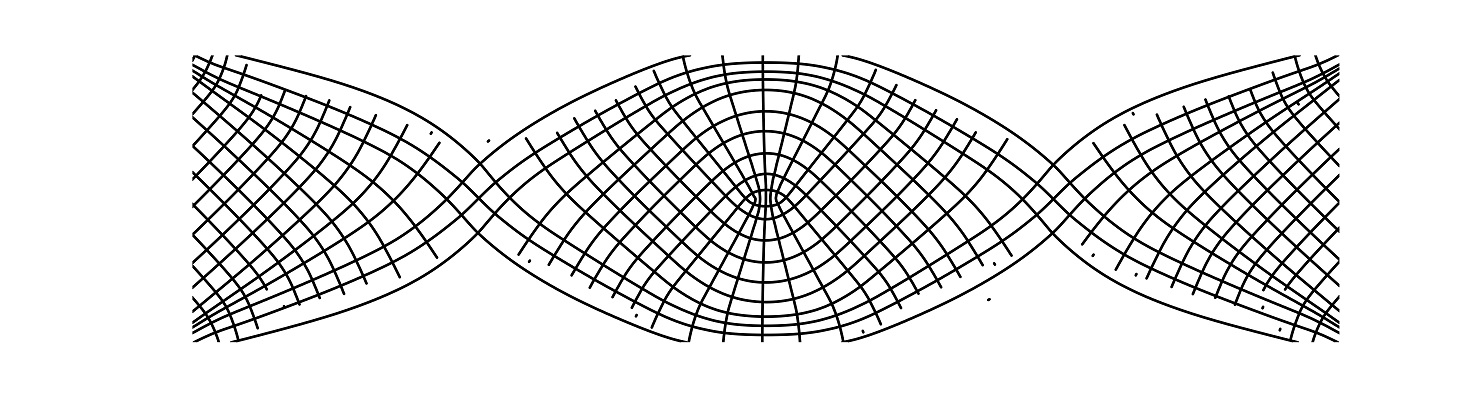}
        \subcaption{The curves selected by the optimizer when using an L1 norm.}
        \label{fig:SubSelection:L1L2Norm:L1Norm:SelectedCurves}
    \end{minipage}
    ~
    \hfill
    ~   
    \begin{minipage}[t]{0.45\linewidth}
        \centering
        \includegraphics[trim=37mm 10mm 12mm 10mm, clip, width=\linewidth,origin=c]{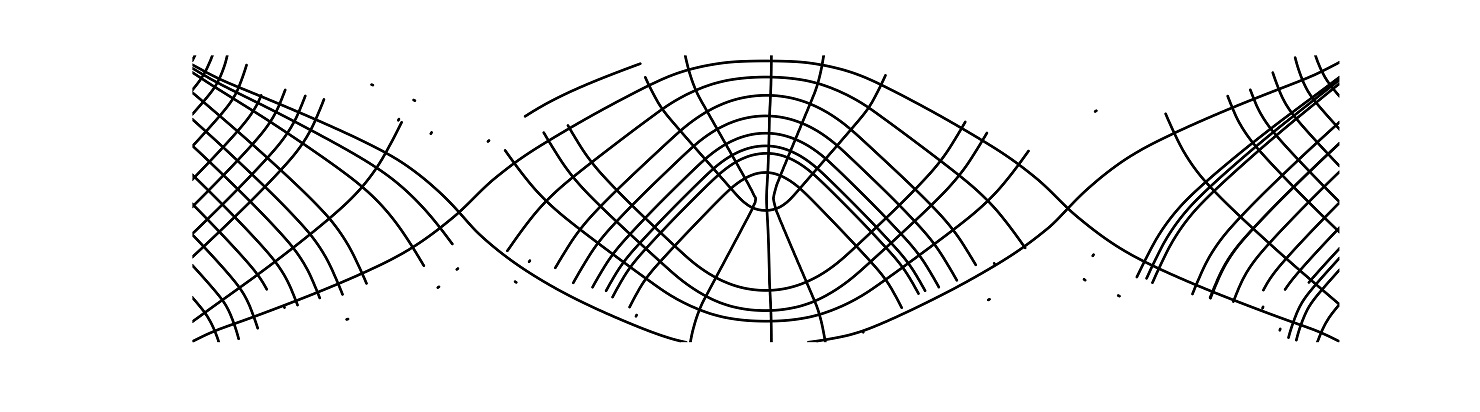}
        \subcaption{The curves selected by the optimizer  when using an L2 norm.}
        \label{fig:SubSelection:L1L2Norm:L2Norm:SelectedCurves}
    \end{minipage}
    \caption{Streamline selection results obtained when providing an L1 norm or an L2 norm to the optimizer. Note the missing streamlines that occur for the L2 norm results due to multiply covered regions being penalized too harsh.}
    \label{fig:SubSelection:L1L2Norm}
\end{figure*}

Note that defining the problem in Equation \ref{eq:SubselectionOptimization} as a least-squares problem instead of a least absolute deviations problem would overly punish multiple covered regions of the energy and lead to missing streamlines. The $L^{1}$ norm in Equation \ref{eq:SubselectionOptimization} on the other hand, penalizes double covered probe points equally as hard as non-covered probe points. The difference of using an L1-norm or an L2-norm can be seen in Figures \ref{fig:SubSelection:L1L2Norm}. Details on the solution of the minimization problem in Equation \ref{eq:SubselectionOptimization}  are discussed in Section \ref{sec:ComputingSubselection:subsec:Optimization}. We now continue to find the variable $n_{\mathcal{S}}$.

\subsection{Number of Streamlines \texorpdfstring{$n_{\mathcal{S}}$}{ns} in the Covering Set of Streamlines \texorpdfstring{$\mathcal{S}$}{S}}
\label{sec:ComputingSubselection:subsec:CardinalityOfStreamsurfaces}

To solve the minimization problem in Equation \ref{eq:SubselectionOptimization} we need to know how large the number of streamlines $n_S = \vert \mathcal{S} \vert$ provided to the optimizer needs to be.

First, we need to define the desired average spacing $\gamma$ of the streamlines in $\mathcal{S}_{opt}$. Then the cardinality of $\mathcal{S}_{opt}$ can be approximated by,
\begin{equation}
    \vert \mathcal{S}_{opt} \vert = \frac{n_{x}}{\gamma} + \frac{n_{y}}{\gamma} \enspace ,
\end{equation}
where $n_{x}$ and $n_{y}$ are the dimensions of the design space in $\mathbf x$, respectively $y$ direction. Note that the cardinality of $\mathcal S_{opt}$ grows in linear dependence to the dimensions of the design space, since streamlines are one dimensional objects. This means that doubling all dimensions of the design space will only lead to a doubling of the cardinality of $\mathcal S_{opt}$. This also holds true in three dimensions, here due to the two-dimensionality of stream surfaces.
We further need to define the error $\varepsilon$ by which a streamline should deviate on average from its optimal position. We denote this in fraction of the optimal average spacing $\gamma$, e.g. $\varepsilon = 0.1$ would allow a streamline to be placed in a band of $0.2 \gamma$ width around its optimal location.
We can then derive the cardinality of $\mathcal{S}$ by:

\begin{equation}
    n_{\mathcal{S}} = \vert \mathcal{S} \vert = \frac{1}{\varepsilon} \vert \mathcal{S}_{opt} \vert = \frac{1}{\varepsilon} \left(\frac{n_{x}}{\gamma} + \frac{n_{y}}{\gamma} \right) \enspace .
\end{equation}
As with $\mathcal{S}_{opt}$, we note that the cardinality of $\mathcal{S}$ grows linear with the dimensions of the design space. We also note that the cardinality of $\mathcal{S}$ grows linear in dependence to the desired error $\varepsilon$, meaning that reducing $\varepsilon$ by a factor $k$ will increase the cardinality of $\mathcal{S}$ only by a factor $k$. Both these observations are again valid in two dimensions as well as in three dimensions.

We have now computed how many stream surfaces we need to provide to the minimization problem in Equation \ref{eq:SubselectionOptimization} to obtain good results.

\subsection{Resolution of the Energy}
\label{sec:ComputingSubselection:subsec:ResolutionOfEnergy}
To solve the minimization problem in Equation \ref{eq:SubselectionOptimization} the only thing that remains is to discretize the energy $E_{S}$ on a pixel grid, where we refer to a single pixel as a probe point. To efficiently subselect streamlines, we need to know the resolutions of the discretized energies, i.e. the number of probe points needed to differentiate streamlines in the set $\mathcal{S}$. This number depends on the desired error $\varepsilon$ and the desired average spacing $\gamma$.
Each streamline should activate the probe points lying in a band of width $\gamma$ around the streamline. Two streamlines that are more than $\varepsilon \cdot \gamma$ apart should activate a different set of probe points. This implies that the number of probe points needed can be computed by,
\begin{equation}
    n_{p} = \frac{n_{x}}{\varepsilon \cdot \gamma} \cdot \frac{n_{y}}{\varepsilon\cdot \gamma} =  \frac{1}{\varepsilon^{2}} \left( \frac{n_{x}}{\gamma} \cdot \frac{n_{y}}{\gamma} \right) \enspace .
\end{equation}
Here we note that the number of probe points grows quadratically in two dimensions, meaning doubling both dimensions of the design space will increase the number of probe points needed by a factor of four. Respectively, the number of probe points grows cubically in three dimensions. Note, however, that the subselection is only a fraction of the time spent on the whole approach as can be extracted from \autoref{tab:timings}.

We have now discretized the energy $E_{S}$  in the minimization problem in Equation \ref{eq:SubselectionOptimization} and are now ready to solve it.

\subsection{Subselection using a Relaxed Approach to Binary Programming}
\label{sec:ComputingSubselection:subsec:Optimization}

Solving the minimization problem in Equation \ref{eq:SubselectionOptimization} can be done by using integer linear programming. However, the underlying problem is likely NP-hard due to the binary constraints. This makes a direct solve of the problem formulated in Equation \ref{eq:SubselectionOptimization} infeasible. To solve the least absolute deviations problem, we relax the optimization variables $w_{i}$ to be in the interval $[0,1]$ instead of $\{0,1\}$. This leads to the following convex linear program, which can be solved in polynomial time:
\begin{equation}
    \label{eq:SubselectionOptimizationRelaxed}
    \begin{aligned}
        \underset{\mathbf{w} \in [0,1]^{n_{\mathcal{S}}}}{\text{minimize}}
        \int_{\Omega} \left| \sum_{i = 1}^{n_{S}} w(i) E_S(\mathbf{x}) - (1,1) \right| \quad \mathrm{d}\Omega \enspace.
    \end{aligned}
\end{equation}
We solve the relaxed problem in Equation \ref{eq:SubselectionOptimizationRelaxed} with an interior point method and then fix weights that have been set to either $0$ or $1$. Subsequently, we solve a binary program with the remaining weights (typically < 5\% of the original weights) using a branch and cut algorithm. A branch and cut algorithm splits the original problem into sub-problems and uses cutting planes to cut away parts of the possible solution space until an optimal integer solution is found for a sub-problem. If that solution is better than a relaxed solution of a second sub-problem, the second sub-problem does not need to be solved. This is done iteratively until the algorithm converges. For details, we refer to \citet{Padberg1991}. We use the implementation provided in CVX \citep{cvx}.

Note that the high number of binary weights chosen in the relaxed problem is due to the energy having binary values. If we were to base the energy on a signed distance function instead, we would almost exclusively receive non-binary weights as a result from the relaxed problem in Equation \ref{eq:SubselectionOptimizationRelaxed} since the optimizer would try to trade off contributions of different streamlines.

The observation in \autoref{sec:ComputingSubselection:subsec:CardinalityOfStreamsurfaces} that the computational burden of the problem in Equation \ref{eq:SubselectionOptimization} grows linear in the amount of stream surfaces has an important practical use. Forking stream surfaces, which  can occur due to heavy noise in the topology optimized fields and are described in \autoref{sec:singularity}, will cover more space than non-splitting surfaces. They are therefore chosen less by the optimizer when the number of surfaces in $\mathcal{S}$ increases.

We have now found a well-spaced set of laminar surfaces $\mathcal{S}_{opt}$ and can now continue with the generation of output structures.

%% file: Sections/00StructureGeneration.tex
\section{Output Generation}
\label{sec:OutputGeneration}
The stream surface tracing and selection procedure described above produces a set of stream surfaces, $\mathcal{S}_{opt}$, each represented as a point cloud. In itself, this representation is useful for visualization. However, our end goal is to provide methods for synthesizing output structures. Here we present a method that produces a volumetric solid from the stream surfaces by compositing a small implicit primitive into a voxel grid for each point in the point cloud representing the stream surfaces. As an additional output modality, we also describe a mesh generator that converts stream surfaces to hexahedral meshes. While imposing restrictions on the proximity of the stream surfaces, the mesh generator can mesh certain non-integrable frame fields.
\subsection{Post-Processing the Surfaces}
\label{sec:super_sample}

When constructing the initial set of surfaces, we do not need a particularly high density of points in each surface. We only need enough to be able to compute the activation of probe points. However, a high density of points will provide a smoother and more precise volumetric solid and will make it easier to tell surfaces apart when computing a hex mesh.

It is crucial that our post-processing still adheres to the field and follows the initial surface. Therefore, our up-sampling is a continuation of the generation procedure. We initialize a new grid for our Poisson Disk Sampling procedure with a much smaller allowed point distance. We then add all of the original points of a selected stream surface to the grid. Subsequently, we add all original points to a queue and restart the point generation. This way, we fill in additional points between the original points since we have chosen a smaller distance for the Poisson Disk Sampling. Any new position is generated as the average of estimates from neighboring points, hence the super-sampled surface will still be following the field closely.

%

\subsection{Volumetric Solids}
\label{sec:volumetric-solids}
A volume representation is convenient for shapes of complex topology, and we can efficiently synthesize a volumetric representation from a stream surface collection.

We compute the volume representation of each stream surface as a sum of  basis functions. This is very similar to methods used for volumetric reconstruction of surfaces from point clouds, e.g. \cite{Fuhrmann14}, except that we design the basis functions to attain their maximum value at the origin. In contrast, in point cloud reconstruction, the basis function's zero level contour typically passes through the origin. The specific basis function that we employ is,
\newcommand{\smoothstep}{\mathrm{ss}}
\begin{equation}
    \phi_i(\mathbf x) = \smoothstep(-\tau_i,0,- |\mathbf n_i \cdot (\mathbf x - \mathbf p_i)|)    \enspace ,
\end{equation}
where $\mathbf p_i$ is the position of point $i$ which is orientated according to the normal, $\mathbf n_i$, and has thickness $\tau_i$. Finally, $\smoothstep$ is the \textit{smoothstep} function,
\begin{equation}
    \begin{aligned}
         & \smoothstep(a,b,x) = 3 t^2 - 2 t^3,                                                  \\
         & \quad \mathrm{where,} \quad t= \min\left(1,\max\left(0,\frac{x-a}{b-a}\right)\right)
        \enspace .
    \end{aligned}
\end{equation}
For each stream surface, $s$, we compute the volumetric representation as a sum for each voxel,
\begin{equation}
    V_s[\mathbf x] = \frac{\sum_i w_i(\mathbf x) \phi_i(\mathbf x)}{\sum_i w_i(\mathbf x)} \enspace ,
    \label{eq:stream-surface-splatting}
\end{equation}
where $V_s$ is the voxel grid, $\mathbf x$ ranges over the positions of all voxels, $s$ indexes the stream surface, and the weight, $w_i$, is given by
\begin{equation}
    w_i(\mathbf x) = \smoothstep(-r, 0, - \|\mathbf x - \mathbf p_i - \mathbf n (\mathbf n_i \cdot (\mathbf x - \mathbf p_i))\|) \enspace .
\end{equation}
The product of $w_i$ and $\phi_i$ is non-zero only in a cylindrical region of radius $r$ centered in point $i$. We only need to consider this region when adding the contribution to \autoref{eq:stream-surface-splatting}. The weights are summed to a separate grid, and then normalization is performed in a second step.

Having computed a volumetric representation of each stream surface, the volumetric solid corresponding to stream surface collection is simply the union of the solids for each of the stream surfaces. The union is computed as the maximum over all stream surfaces,
\begin{equation}
    V[\mathbf x] = \max_s(V_s[\mathbf x]) \enspace.
\end{equation}

Finally, we compute a triangle mesh of the boundary of the volumetric solid using  dual contouring with the iso-value 0.5 \cite{Ju02}.
\subsection{Hexahedral Meshes}
\label{sec:hexes}
For certain frame fields, we are also able to produce hexahedral meshes. For reasons explained in \autoref{sec:stc}, we require a minimum separation between stream surfaces of $4r$ where $r$ is the minimum distance between samples on a single stream surface. In practice, this requirement is not possible to fulfill for our topology optimization examples, but with this approach, we can, in fact, handle certain non-integrable fields.

\begin{figure}[h]
    \centering
    \includegraphics[width=0.8\columnwidth]{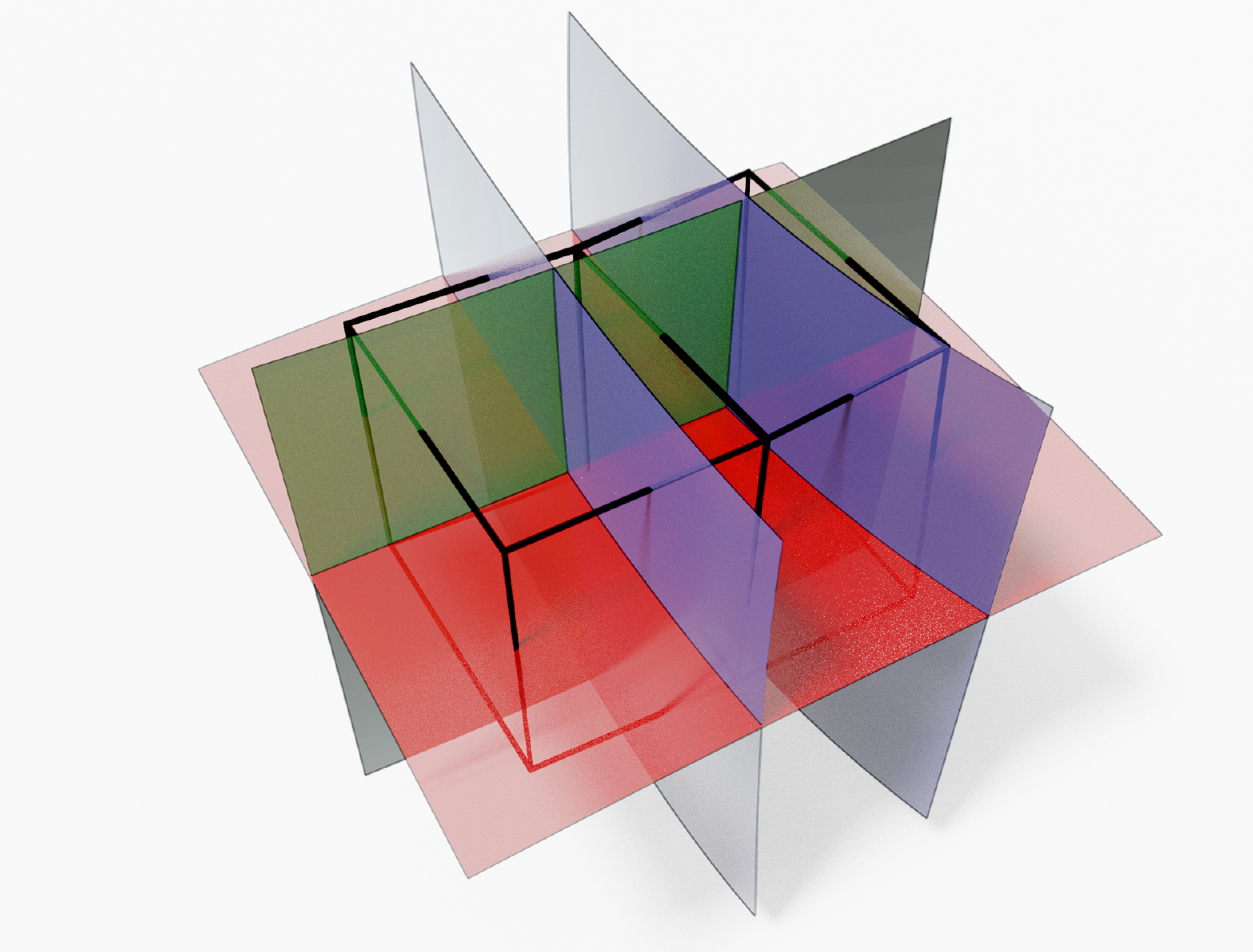}
    \caption{The two hexahedra (dark wireframe) are dual to the structure formed by intersecting the four surfaces.}
    \label{fig:stc}
\end{figure}

The hex meshing approach is based on the following observation. Given a collection of surfaces in 3D, we can form a curve network by finding all the intersecting curves between pairs of surfaces.
It has been observed by \citet{Murdoch97} that such a system of interlocking surfaces, called  the \textit{spatial twist continuum} (STC), is the dual of a hexahedral mesh. It is relatively easy to see that this is true if we assume that surfaces only intersect along a curve and that no more than three surfaces meet at a single point. Since in that case, all the vertices formed as triple intersections must have valence six and thus correspond to a hexahedral cell in the dual. A very simple example with two hexahedra formed by dualizing an STC consisting of four surfaces is shown in Figure~\ref{fig:stc}.
\subsubsection{Constructing the STC Graph}
\label{sec:stc}
Since our stream surfaces are represented as point clouds, we cannot directly compute the intersections. Instead, we formulate the problem as a graph problem. We form a graph, $G = \left<P,E\right>$, whose vertices, $P$, are the union of the points in all stream surfaces, i.e. $P=\bigcup_j S_j \in \mathcal{S}_{opt}$. Two vertices are connected by an edge in $E$ if their distance is smaller than $2r$, where $r$ is the distance used in the stream surface super-sampling. Assuming that the stream surfaces corresponding to the same lamination direction are always further away than $4r$, no vertex should have neighbors belonging to more than three stream surfaces. In fact, the vertices of $P$ belong to three classes: \textit{surface vertices} all of whose neighbors belong to a single stream surface, \textit{intersection vertices} whose neighbors belong to two stream surface, and \textit{triple intersection vertices} whose neighbors belong to three stream surfaces.

Forming connected components of vertices which satisfy the equivalence relation, we can now create a model of the STC as a new graph $G_\mathrm{STC} = \left<P_\mathrm{STC}, E_\mathrm{STC}\right>$ from $G$ based on the vertex classification. Initially, we discard all surface vertices and form the vertex set, $P_\mathrm{STC}$, by creating a vertex for each connected component of triple intersection vertices of $P$. The vertex connectivity is found using a simple run of Dijkstra's algorithm on $G$. We initialize all triple intersection vertices with distance 0 and compute the graph distance to all intersection vertices. Each such vertex is also assigned the index of its predecessor, allowing us to recursively trace back to the originating cluster. Once all intersection vertices have been visited, we visit all edges in $G$. If the two incident vertices were reached from different clusters, then the vertices in $P_\mathrm{STC}$ that correspond to these clusters are connected by an edge in $E_\mathrm{STC}$.
\subsubsection{Hexahedralization}
From  $G_\mathrm{STC}$ we compute a hexhedral mesh by computing the dual. A hexahedron is created for each vertex in $P_\mathrm{STC}$. The hexahedron is scaled to the average length of edges incident on the vertex and rotated to align with the same edges. Finally, we assign the best-aligned quadrilateral face with each outgoing edge of the vertex.

To construct the connectivity of the hexahedral mesh, we now visit all edges in $E_\mathrm{STC}$ and cluster the vertices of the associated quads. Since every vertex is associated with up to eight hexahedra, these clusters may be of size up to eight. We compute each cluster's barycenter and assign this as the position of the vertex in the final hex mesh.

This method can only produce hexahedra. Hence, the frame field singularities do not result in irregular cells but gaps in the mesh or irregular vertices.  The former is shown in \autoref{fig:helix-hexes} where a field that spirals around a central axis has been hexahedralized. Since stream surface tracing stops at the central singularity, it leaves a gap in the mesh. The latter effect is shown in \autoref{fig:sphere-hexes} where the hexahedra can be observed to have slightly worse quality near the network of singular curves inside the sphere.

%% file: Sections/00Results.tex
\section{Implementation and Results}
\label{sec:Results}

\label{sec:Results:Implementation and Performance}
Our implementations are in C++ and Matlab. Codes related to the creation and manipulation of surfaces are primarily written in C++. Codes for selecting surfaces are running in Matlab by use of the CVX package \cite{cvx,Grant2008} and the Mosek solvers \cite{Mosek}.

The generation of stream surfaces and the synthesis of volumetric solids are parallel processes and have been parallelized using MPI and native threading facilities of C++, respectively. The stream surface tracing and the subselection were executed on a node equipped with two Intel Xeon E5-2650 v4 processors. The volumetric solid and hex meshing was executed on a single Intel Core i7. An overview of the statistics, including computation time, is shown in \autoref{tab:timings}.

\begin{table*}[ht]
    \centering
    \caption{Here we show the statistics for the different steps in our pipeline. The first three blocks of rows show relevant statistics for the initial point sampling of stream surfaces, the selection of optimal stream surfaces, and the super-sampling of the selected stream surfaces. The fourth block reports statistics for generating the volumetric solids and the fifth block shows the time to run the hexahedralization algorithm. We then report in the sixth block the overall runtime of our approach. We include the times for the homogenization-based topology optimization and report them in the last block for completeness. Note how our algorithm is only a fraction of the topology optimization runtime.}
    \label{tab:timings}

    \begin{adjustbox}{width=\textwidth}
        \begin{tabular}{p{4.5cm}|ccccccc}
            \hline
                                                          & \multicolumn{1}{c}{\begin{tabular}[c]{@{}c@{}}Cantilever\\ (3 layers)\end{tabular}} &
            \multicolumn{1}{c}{\begin{tabular}[c]{@{}c@{}}Cantilever\\ (1 layer)\end{tabular}} &
            \multicolumn{1}{c}{\begin{tabular}[c]{@{}c@{}}Electrical Mast\\ \  \end{tabular}} &
            \multicolumn{1}{c}{\begin{tabular}[c]{@{}c@{}}Torsion Sphere\\ \  \end{tabular}} &
            \multicolumn{1}{c}{\begin{tabular}[c]{@{}c@{}}Sphere\\ \  \end{tabular}} &
            \multicolumn{1}{c}{\begin{tabular}[c]{@{}c@{}}Helix \\ \  \end{tabular}} &
            \multicolumn{1}{c}{\begin{tabular}[c]{@{}c@{}}Cylinder \\ \  \end{tabular}}                                                                                                                                                                        \\

            \hline
            $|\mathcal{S}|$ (Number of stream surfaces)   & 480                                           & 478                     & 450                     & 374                     & 480          & 170        & 400        \\
            Generating $\mathcal{S}$ (runtime)            & 20 min                                        & 28 min                  & 17 min                  & 1h 51 min               & 49 min       & 1 h 4 min  & 1 h 1 min  \\
            Average points per surface                    & 961                                           & 1 351                   & 667                     & 5 043                   & 2 019        & 5 082      & 1 906      \\
            \hline
            Subselection (runtime)                        & 45 sec                                        & 53 sec                  & 2 min 5 sec             & 2 min 31 sec            & 39 sec       & 50 sec     & 37 sec     \\
            $|\mathcal{S}_{opt}|$                         & 38                                            & 22                      & 35                      & 4                       & 27           & 24         & 41         \\
            \hline
            Super-sampling (runtime)                      & 2 h 30 min                                    & 3 h 18 min              & 1 h 19 min              & 1 h 30 min              & 1 h 52 min   & 2 h 9 min  & 6 h 20 min \\
            Average points per
            \newline super-sampled surface
                                                          & 9448                                          & 9 735                   & 3 483                   & 58 148                  & 25 073       & 26 320     & 15 535     \\
            \hline
            Output volume dimension                       & $700\times350\times350$                       & $700\times350\times350$ & $266\times266\times800$ & $500\times500\times500$ &              &            &            \\
            Solid generation (runtime)                    & 13 min 42 sec                                 & 11 min 3 sec            & 5 min 48 sec            & 8 min 30 sec            &              &            &            \\
            \hline
            Hexahedralization (runtime)                   &                                               &                         &                         &                         & 1 min 24 sec & 23 sec     & 38 sec     \\
            \hline
            Summed runtime                                & 3 h 04 min                                    & 3 h 58 min              & 1 h 44 min              & 3 h 32 min              & 2 h 43 min   & 3 h 14 min & 7 h 20 min \\
            \hline
            \hline
            Topology optimization time                    & 7 h 48 min                                    & 9 h 54 min              & 22 h 20 min             & 40 h 21 min                                                      \\
            \hline
        \end{tabular}
    \end{adjustbox}
\end{table*}

\subsection{Missing Structural Members and Field Alignment -- a Comparison}
\begin{figure}[ht]
    \centering
    \begin{minipage}[t]{0.95\columnwidth}
        \centering
        \includegraphics[trim=0mm 0mm 0mm 0mm, clip, width=\textwidth,origin=c]{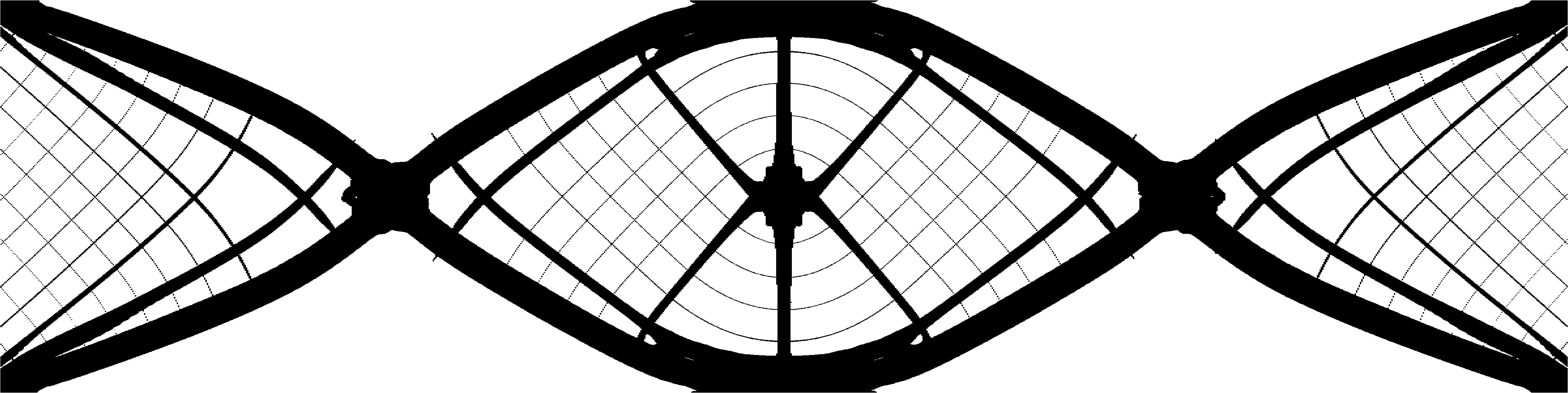}
        \subcaption{De-homogenization using an integrative approach proposed in \citet{Stutz2020} yielding missing structural members around the singularity, where the gradients have become almost zero. A resolution of $320\times80$ is used as an input mesh. The shown Figure has a compliance-volume value of $C \cdot V =  7.30$. \citet{Stutz2020} report Compliance-Volume values of 7.05, 7.48, 7.63 and 21.39 for different alignment weights, where 7.05 is their best performing structure at an intermediate alignment weight and 21.39 is a failure case.}
        \label{fig:2DMissingStructuralMembers:Integrative}
    \end{minipage}
    ~
    \hfill
    ~
    \begin{minipage}[t]{0.95\columnwidth}
        \centering
        \includegraphics[trim=0mm 0mm 0mm 0mm, clip, width=\textwidth,origin=c]{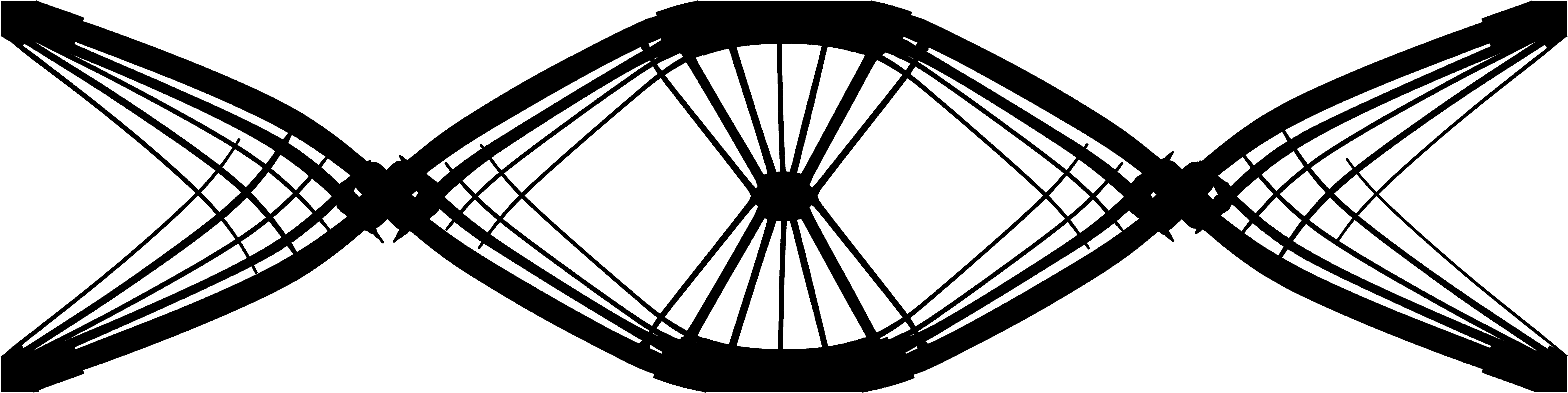}
        \subcaption{Our approach yields a structure with evenly spaced structural members for the same $320\times80$ input field as used above. The compliance-volume value for this structure is $C \cdot V = 7.44$ and is therefore a mere 5.5\% worse than \citet{Stutz2020} best value. However, with our approach we do not risk a failure case.}
        \label{fig:2DMissingStructuralMembers:Streamlines}
    \end{minipage}
    \caption{Comparison between an integrative approach based on \citet{Stutz2020} yielding missing structural members and our approach which creates evenly spaced structural members.}
    \label{fig:2DMissingStructuralMembers}
\end{figure}

As discussed in \autoref{sec:RelatedWork} our method aims to circumvent the problem of missing structural members due to enforcement of alignment to the input field when using an integrative method to create a parametrization. As discussed in \citet{Groen2018,Groen2020, Stutz2020}, alignment of the final structure to the input field needs to be enforced by a constraint when adapting integrative approaches as \citet{Kaelberer2007, Bommes2009, Nieser2011}. This is done by enforcing the parametrization to be orthogonal to the second (and third) normal direction. However, if this alignment is too strict, the gradient of the parametrization may become almost zero in large regions. This, in turn, can then lead to overly thick structural members or to missing structural members especially around singularities as discussed by \citet{Stutz2020}. Our approach creates well-aligned structures before selecting a subset, eradicating the problem since we cannot suffer from vanishing gradients, as we do not integrate the fields. We show an example in \autoref{fig:2DMissingStructuralMembers}. Note that the same behavior can be observed in three dimensions.

The structure shown in \autoref{fig:2DMissingStructuralMembers:Integrative} has been obtained dehomogenizing a $320\times80$ layer-normal field by an integrative approach proposed by \citet{Stutz2020}. Note how there are missing structural members above and underneath the singularity. The structure has a compliance $C = 26.46$ and a volume fraction of $V = 0.275$. For comparison we use the compliance-volume value $C \cdot V = 7.30$. \citet{Stutz2020} report compliance-volume values of 7.05, 7.48, 7.63, and 21.39 for different alignment weights at the same resolution. Here 7.05 is their best performing structure at an intermediate alignment weight, and 21.39 is a failure case.

\autoref{fig:2DMissingStructuralMembers:Streamlines} shows the structure created by our approach using also a $320\times80$ layer-normal field as was used by \citet{Stutz2020} in \autoref{fig:2DMissingStructuralMembers:Integrative}. Note that our approach yields a structure with evenly spaced structural members. The structure has a compliance $C = 27.67$ and a volume fraction of $V = 0.269$. The compliance-volume value for this structure is $C \cdot V = 7.44$. Note that this value is only 5.5\% worse than \citet{Stutz2020} best value. Moreover, with our approach we do not risk a failure case due to bad alignment or zero gradients in a parametrization.

\subsection{Volumetric Structures from Topology Optimized Fields}
\label{sec:Results:VolumetricStructures}

\begin{figure*}[ht]
    \includegraphics[width=0.98\linewidth]{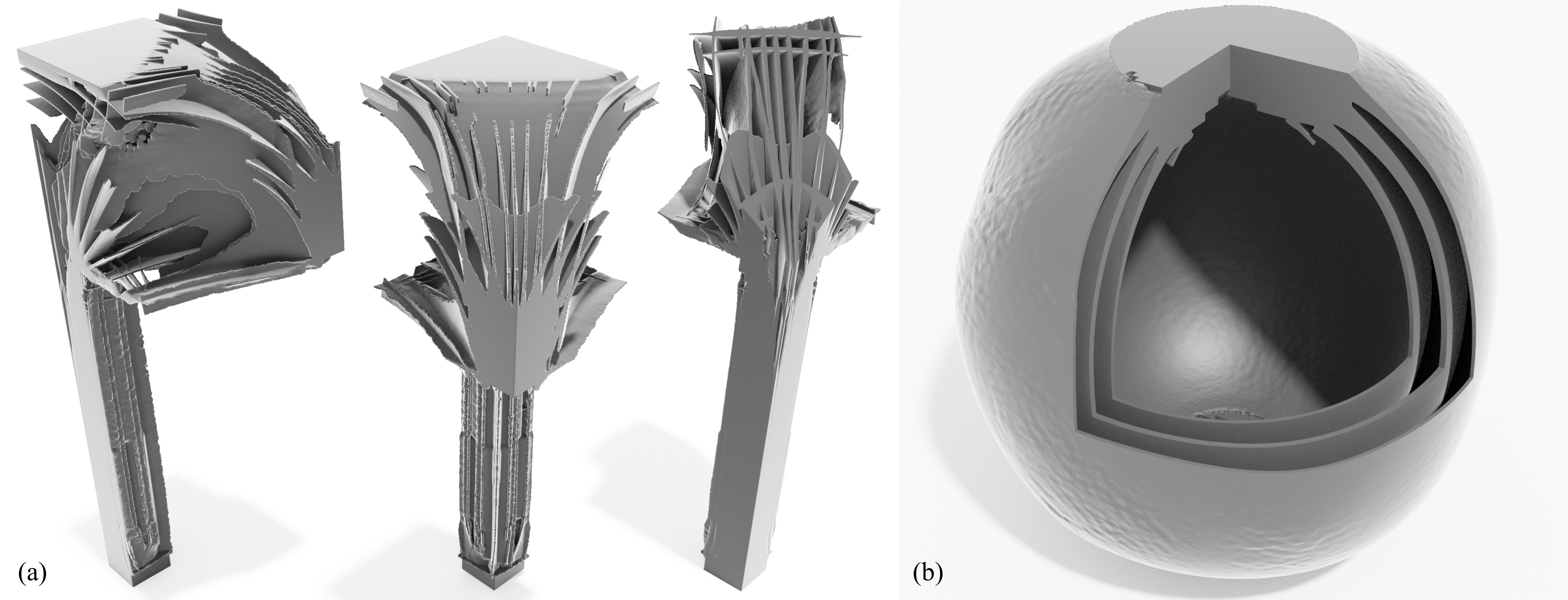}
    \caption{This figure show two topology optimization examples created with  our method from \autoref{sec:volumetric-solids}. In (a) we see the electrical mast example introduced in \citet{GeoAllOli20}. We show here three angles (side, front and back), where in the rightmost image showing the mast from the back, the top has been cut away to reveal the interior of the structure.
        In (b) we show the Michell torsion sphere example introduced in \citet{Groen2020}, where the boundary conditions with torsion applied are located in the top and the botton. Note that we cut out an eigth to reveal the interior laminations of the torsion sphere.}
    \label{fig:splatting-dehomogenization}
\end{figure*}

\begin{figure*}[ht]
    \includegraphics[width=0.98\linewidth]{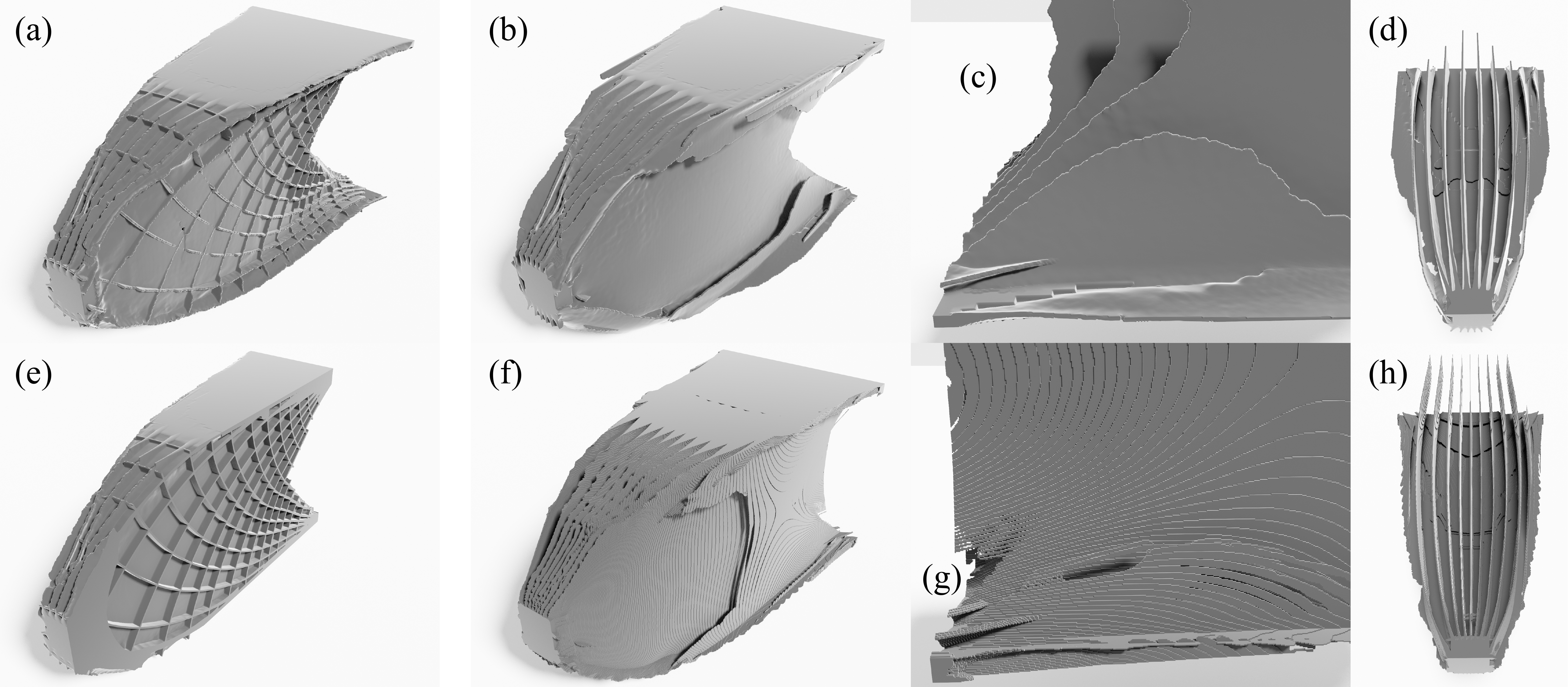}
    \caption{On the left, (a,e) show a cantilever produced from a homogenization solution where all three layers have been forced to be active. The images on the right show a comparison of our results for the cantilever (b,c,d) with the results produced using the method of \citet{Groen2020} (f,g,h).}
    \label{fig:cantilever-dehomogenization}
\end{figure*}

We ran various input fields from topology optimization through our pipeline. The fields were generated by the method proposed by \citet{Groen2020}. For problem formulations of the homogenization-based topology optimization and a description of the load cases, we refer to \citet{GeoAllOli20, Groen2020}. The timings of the field generation and the de-homogenization are reported in \autoref{tab:timings}, where we see that the topology optimization dominates over our de-homogenization approach. In \autoref{fig:splatting-dehomogenization}a we see a quarter of an electrical mast as proposed in \cite{GeoAllOli20}. The fields generated for the electrical mast example contain spurious singularities in fully solid regions and the void due to the microstructure being isotropic (solid) or non-existent (void).
Nevertheless, we produce very smooth surfaces, since our streamsurfaces do not need to expand into solid or void regions. \citet{Groen2020} make use of the fact that singularities only arise in fully solid or void regions by combing the fields in intermediate regions first, such that the spurious singularities cannot create seams in the combed field that extend into the intermediate regions. However, their approach yields no control or guarantee over how much singularities influence their designs since they still rely on the orientations in solid and void for integration, although they use relaxation for such elements.

On the right, in \autoref{fig:splatting-dehomogenization} we show our version of the torsion sphere example proposed in \citet{Groen2020} that was based on Michell's famous torsion sphere \cite{M.C.E.1904}. Note that since we use optimal rank-3 microstructures, we do not get a truss structure, but a stiffer layer structure \cite{Sigmund2016}. The torsion sphere has a singularity that connects the two boundary conditions, similar to a towel being wrung out.
This singular curve passes through the solid region at one boundary condition, then through the void and the solid region at the opposite boundary condition. Note that this is again not a problem for our algorithm since we neither need to expand into void nor solid regions. \citet{Groen2018} include the singular curve in their integration of the field without any special measurement, since they are relaxing the parametrization in the void and fully solid regions. Our method produces three high-quality shells that align well with the input field.

\autoref{fig:cantilever-dehomogenization} shows the three dimensional version of Michell's cantilever. For loading cases and problem formulation we refer to \citet{GeoAllOli20, Groen2020}. We compute de-homogenization results for two versions. In \autoref{fig:cantilever-dehomogenization}a, we depict a solution for the cantilever where we enforce that either all three layers have a layer thickness of more than 5\% or that all layer thicknesses are zero. Such a design is great for resistance against buckling. Note that due to all three layers being enforced to have non-zero layer widths outside of the void, the microstructure orientation becomes unique in this example. Spurious singularities only arise in solid and void regions; therefore, we have no problems with forking stream surfaces in this example neither. A cut section through the structure is given in \autoref{fig:cantilever-dehomogenization}e.

In \autoref{fig:cantilever-dehomogenization}b, we show the second version of the cantilever that we consider. These input fields have been created without any enforcement on the layer-thicknesses and correspond to the cantilever \citet{Groen2020} propose. We compare our results with theirs, first on a visual level in Figures \ref{fig:cantilever-dehomogenization}b, c, d, f, g and h and then in terms of compliance and volume in Table \ref{tab:cantileverCompliance}.

In Figures \ref{fig:cantilever-dehomogenization}b and f we see the full de-homogenized structures. The two structures are very alike. Note that \citet{Groen2020} use some additional fine-scale evaluation to remove unused excess material, i.e.\ low strain energy elements. This puts their structure at a slight advantage over ours, since we do not incorporate such a step for our structure in Figures \ref{fig:cantilever-dehomogenization}b-d. Figures \ref{fig:cantilever-dehomogenization}c and g show a detail and Figures \ref{fig:cantilever-dehomogenization}d and h show horizontal cuts through the structures.

Since our input fields differ from \citet{Groen2020} we cannot compare the values in Table \ref{tab:cantileverCompliance} with too much emphasis. Note also that \citet{Groen2020} evaluated their design using a fine scale $960\times480\times480= 221\ 184\ 000$ finite elements model, where as we only evaluate our model on $512\times256\times256 = 33\ 554\ 432$ elements. We therefore also only compare our result to the best performing values that \citet{Groen2020} report for a $96\times48\times48$ frame field as we use. The most meaningful value is certainly $\frac{C_{s} \cdot V_{s}}{C_h \cdot V_{h}}$ which sets the compliance-volume value of the de-homogenized structure to the compliance-volume fraction reported by the topology optimization. We see that we are a mere 5.5\% off the \citet{Groen2020} best performing structure, even though we do no parameter study to find the best performing structure for the de-homogenization parameters since this would be outside of the scope of this paper.

\begin{table}[ht]
    \caption{Comparison of our results for the cantilever example with results obtained by \citet{Groen2020}.
        We use the following abbreviations:
        $V_{s}$ = volume of the de-homogenized structure, $C_{s}$ = compliance of de-homogenized structure, $V_{h}$ = volume of the homogenization-based topology optimization solution, $C_{h}$ = compliance of the homogenization-based topology optimization solution
        Note that \citet{Groen2020} evaluated their design on a $960\times480\times480$ finite elements model, where as we only evaluate our model on $512\times256\times256$ elements.}

    \begin{tabular}{lll}
        \hline
        Cantilever                                  & \citet{Groen2020} & Our approach \\ \hline
        $C_{h}$                                     & 226.68            & 228.45       \\ \hline
        $V_{h}$                                     & 0.1000            & 0.1000       \\ \hline
        $C_{s}$                                     & 243.31            & 223.72       \\ \hline
        $V_{s}$                                     & 0.1021            & 0.1181       \\ \hline
        $ C_{s} \cdot V_{s}$                        & 24.845            & 26.428       \\ \hline
        $\frac{C_{s} \cdot V_{s}}{C_h \cdot V_{h}}$ & 1.0960            & 1.1568       \\ \hline
    \end{tabular}
    \label{tab:cantileverCompliance}
\end{table}
\subsection{Hexahedral Meshes from Boundary Aligned and Closed-Form Fields}
\label{sec:Results:HexMeshes}

Our main goal was to synthesis volumetric structures from frame fields, but it is also possible to generate hex meshes from a collection of stream surfaces. However, it should be noted that while the volumetric synthesis method does not impose demanding requirements on the stream surfaces, hex mesh generation requires a minimum distance between the stream surfaces. Moreover, our topology optimization examples (in particular, the torsion sphere and the cantilever with one active layer) are clearly ill-suited to hex meshing. For this reason, our hex meshing examples are very different from the examples above.

\begin{figure}[ht]
    \includegraphics[width=0.95\columnwidth]{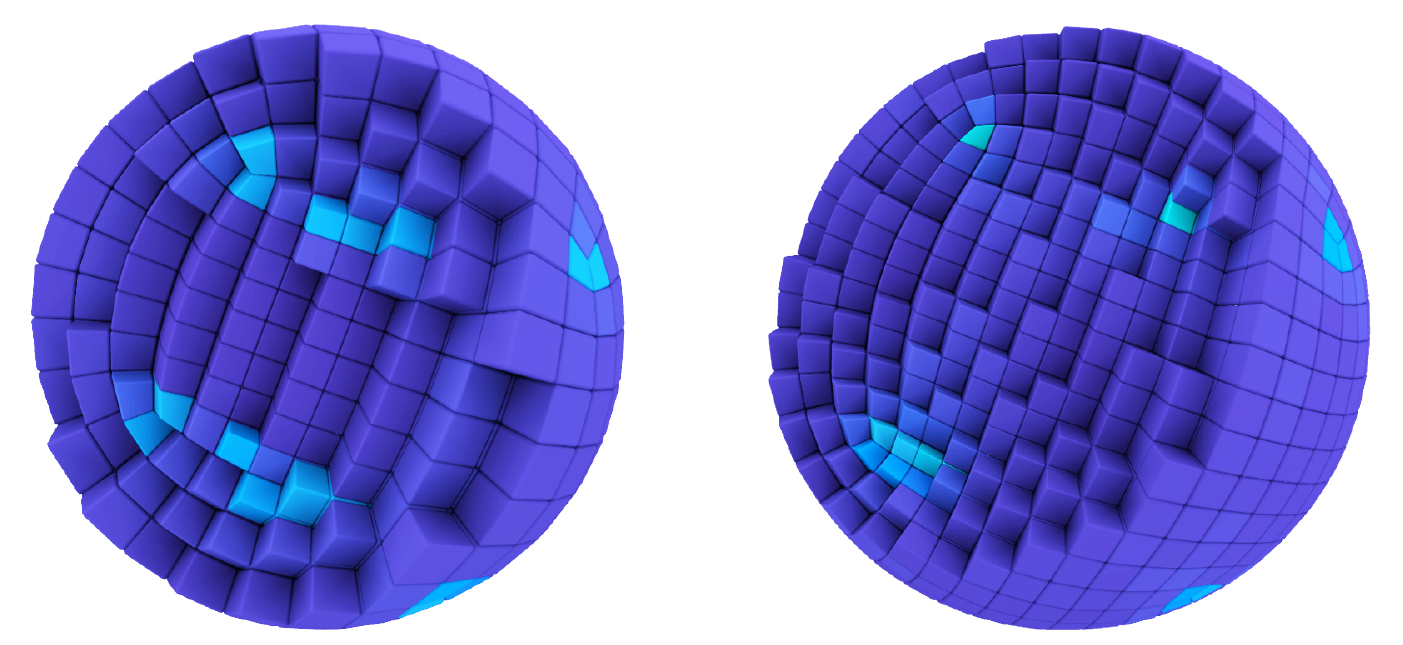}
    \caption{Comparison of a hex mesh of a sphere produced by our method based on an octahedral frame field generated using the method of \citet{Palmer2019} and a similar hexmesh produced from a field created with the method of \citet{Corman2019} and hex meshed using CubeCover in an implementation by David Bommes (right). For both visualization we use \citet{Hexalab}. The hexahedra are colored according to the scaled Jacobian.}
    \label{fig:sphere-hexes}
\end{figure}

\autoref{fig:sphere-hexes} illustrates the hex meshing of a sphere (3328 hexahedra) based on a boundary aligned frame field created using the method of \citet{Palmer2019}. We compared this mesh to one created by \citet{Corman2019} with assistance from David Bommes containing 4032 hexahedra. The meshes are structurally similar, and the min/average scaled Jacobian \cite{Hexalab} for our mesh are 0.425/0.956 vs Corman et al.: 0.474/0.964. Note that since the boundary aligned frame field inside the sphere has been optimized for smoothness, it only contains singularities of index $\nicefrac{1}{4}$. In this specific example our stream surface tracing can be used without cutting out the singular curves, since the field is smooth enough.

\begin{figure}[ht]
    \includegraphics[width=0.95\columnwidth]{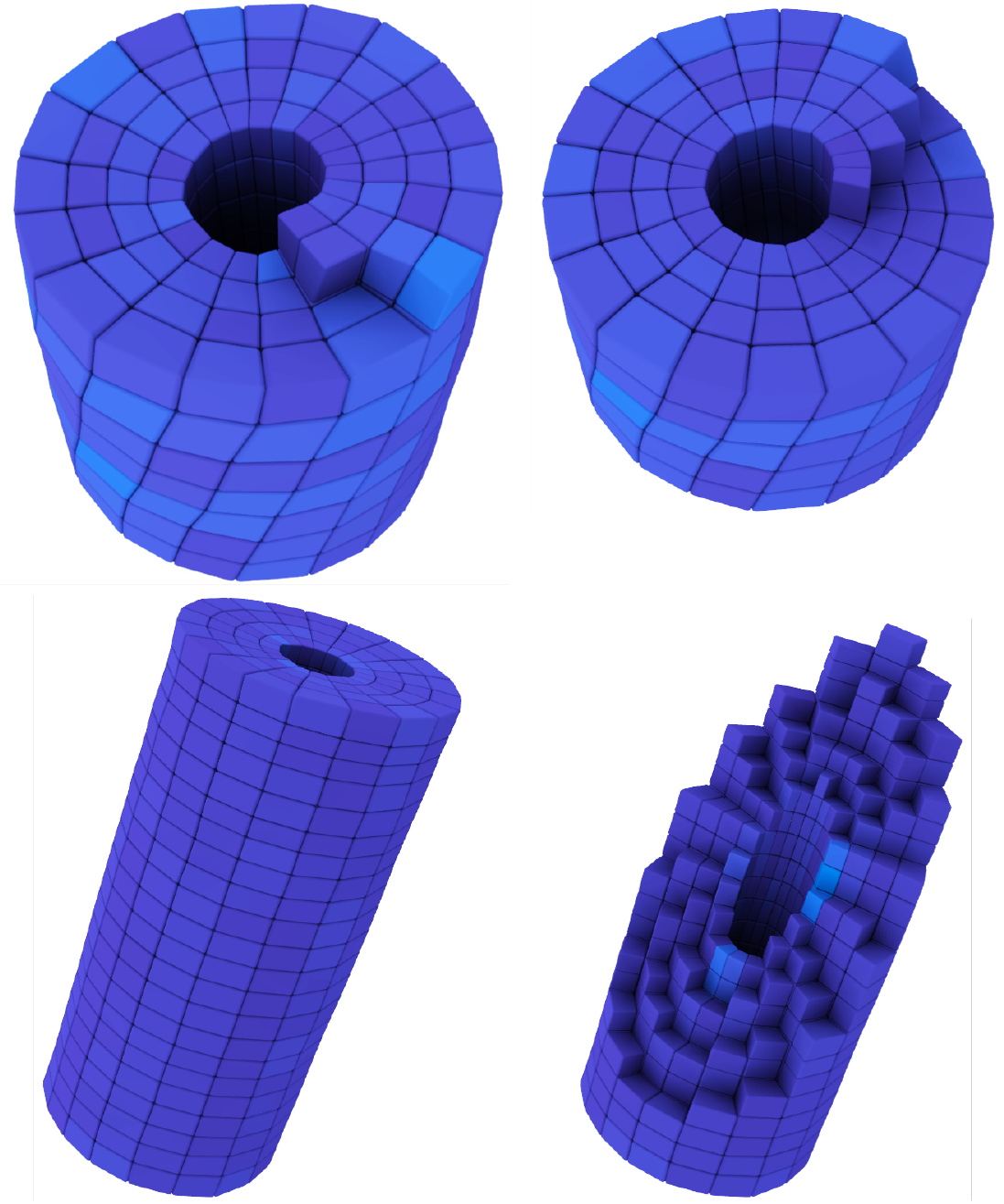}
    \caption{In the top row, we see a hex mesh generated using our method on a field describing a helicoid. Note that the underlying field is non-integrable. On the left, we see the entire mesh, whereas on the right, we have removed several layers of hexahedra.
        In the bottom row, we see a hex mesh generated on a cylindrical field. Note how the shape of the hexahedra changes dramatically from the outside towards the cut out singular curve. On the right, we peel away several layers of hexahedra, which reveals that the minimal edge-length is significantly smaller than the maximal edge-length. Note that we desire this from a field describing an anisotropic mesh.}
    \label{fig:helix-hexes}
\end{figure}

\autoref{fig:helix-hexes} shows hex meshes of a spiral and a cylinder generated based on frame fields from closed-form expressions. In both cases we have an index 1 singularity in the center, and the spiraling frame field is non-integrable making it challenging for many other strategies. The min/average scaled Jacobian are 0.474/0.964 (spiraling frame field) and 0.870/0.977 (cylinder).


%% file: Sections/00Conclusion.tex
\section{Discussion and Future Work}
In this paper, we have introduced a novel method for creating multi-laminar structures that align to frame fields. The main challenge lies in the fact that even though we can easily make a local structure that aligns with the frame field, we cannot easily assemble these into a global structure. One way to approach this is through the introduction of a parametrization of the domain. Indeed, the previous methods of which we are aware require a parametrization of the domain. This is however \textit{only} straightforward to compute in the absence of singularities in the frame field.

While singularities do need to be taken into account with our method, their presence does not fundamentally change the algorithms we use to create and select stream surfaces. In that sense, our approach is (almost) oblivious to singularities. In contrast, a parametrization based approach needs to explicitly deal with singularities by introducing seams and clearly cannot align perfectly to a non-integrable field. For the application of de-homogenization this translates into the pitfall that the parametrization modifies the resulting mechanical structures negatively. Moreover, a practical challenge when computing a parametrization is that the frame field must be combed -- i.e.\ there must be a consistent labeling of the frame vectors. 

\balance

Stream surface tracing and selection sidestep both of these issues, and we have demonstrated that our approach can provide robust output for various types of fields and create highly anisotropic structures and hex meshes outside of the singular region. We do not require any prescribed edge-lengths; the implicit description of anisotropy by the input fields suffices.

Admittedly, there are also several limitations to our approach. Stream surface tracing in highly rotating fields is difficult. We need to stop tracing stream surfaces that cross the singular regions at an angle not perpendicular to the singular curve. However, for topology optimized fields this is no major concern as discussed in Section \ref{sec:inputFields:TopOpt}. We are further able to keep the hole size minimal, such that these regions can mostly be filled with hexahedra for a boundary optimized field if a hexahedral mesh is output. However, our hex meshing scheme is admittedly simplistic. The limit on the proximity of stream surfaces means that we can only reliably generate hexahedral meshes for relatively simple frame fields, whereas there is no restriction on the frame fields for which we can synthesize volumetric solids.
As a further limitation, we have realized that an excess amount of stream surfaces can overload the optimizer. This situation occurs if one field is activated by a large amount of stream surfaces, which are all equally good, such as in the case of the helicoid example (\autoref{fig:helix-hexes}). Here, it is possible to end up with hundreds of helicoidal surfaces each of which activates almost all probe points. Consequently, an important next step is to avoid oversampling  by monitoring which regions (i.e.\ probe points) are already well-covered.

Finally, we have focused on applying the method to fields arising from compliance minimization in this paper, but topology optimization is replete with problems where the proposed method might be applicable, hinting at future application areas.